\documentclass[pra,aps,twocolumn,superscriptaddress]{revtex4-1}
\usepackage{graphicx,color}
\usepackage{amsthm}
\usepackage{amsfonts}
\usepackage{algorithmic}
\usepackage{enumerate}
\usepackage{latexsym}
\usepackage{amsmath}
\usepackage{amssymb}
\usepackage[colorlinks=true,citecolor=blue,linkcolor=blue]{hyperref}

\usepackage{dcolumn}
\usepackage{bm}

\usepackage[T1]{fontenc}
\usepackage{mathptmx}

\begin{document}
\title{Inducing a metal-insulator transition in disordered interacting Dirac fermion systems via an external magnetic field}

\author{Jingyao Meng}
\affiliation{Department of Physics, Beijing Normal University, Beijing 100875, China}
\author{Rubem Mondaini}
\affiliation{Beijing Computational Science Research Center, Beijing 100193, China}

\author{Tianxing Ma}
\email{txma@bnu.edu.cn}
\affiliation{Department of Physics, Beijing Normal University, Beijing 100875, China}
\affiliation{Beijing Computational Science Research Center, Beijing 100193, China}
\author{Hai-Qing Lin}
\affiliation{Beijing Computational Science Research Center, Beijing 100193, China}
\affiliation{Department of Physics, Beijing Normal University, Beijing 100875, China}

\begin{abstract}
We investigate metal-insulator transitions on an interacting two-dimensional Dirac fermion system using the determinant quantum Monte Carlo method. The interplay between Coulomb repulsion, disorder and magnetic fields, drives the otherwise semi-metallic regime to insulating phases exhibiting different characters. In particular, with the focus on the transport mechanisms, we uncover that their combination exhibits dichotomic effects. On the one hand, the critical Zeeman field $B_c$, responsible for triggering  the band-insulating phase due to spin-polarization on the carriers, is largely reduced by the presence of the electronic interaction and quenched disorder. On the other hand, the insertion of a magnetic field induces a more effective localization of the fermions, facilitating the onset of Mott or Anderson insulating phases. Yet these occur at moderate values of $B$, and cannot be explained by the full spin-polarization of the electrons.
\end{abstract}


\maketitle
\section{Introduction}
Since Slater argued that a gap could be opened by magnetic ordering with spin-dependent electronic energy~\cite{RN106}, applying a magnetic field has been a powerful means to elucidate novel phenomena~\cite{10.3389/fspas.2019.00066}. When making the Zeeman field a variable in a system, fascinating properties are induced through interesting physical mechanisms, which have been widely recognized, such as the spin-Hall effect~\cite{PhysRevLett.107.096601}, topological phase transition~\cite{PhysRevB.101.125118}, superconductor-insulator transition in disordered systems~\cite{Nature2007.449.7164}, anomalous Hall states~\cite{Nature2020.5.1}, magnetic ordering transition~\cite{Milat2004,PhysRevB.80.045412}, and metal-insulator transition~\cite{Nature2020.11.1}. Among them, metal-insulator transitions (MITs) in correlated electron systems have long been a central and controversial issue in material science~\cite{RevModPhys.70.1039}. In Si MOSFETs~\cite{PhysRevB.101.045302,PhysRevB.99.155302,RN122,RN4} or graphene~\cite{NaturePhysics.8.7,RN17,RN18}, a magnetic field has been found to suppress metallic behavior giving way to an insulating phase. In real materials, disorder and interactions are both present, and thus, to fully understand these problems, one needs to treat the challenging interplay between the magnetic field, interactions and disorder on the same footing~\cite{RN1,RN3}.

Recently, the physics in Dirac fermion systems with magnetic fields have attracted intensive studies, including on graphene, topological insulators~\cite{RN116} and Weyl semimetals~\cite{RN120}. Graphene is one of the most promising 2D materials~\cite{RN108,RN109} due to its unique characteristics, such as excellent electrochemical performance~\cite{LI2018845} and ultrahigh electrical conductivity~\cite{RN113,RN114}. When subjecting it to a magnetic field, interesting physics can be revealed through the conductivity's behavior. For instance, comparison of the bulk and edge conductance is crucial for understanding symmetry breaking in the quantum Hall effect (QHE)~\cite{RN17,RN18}, or a parallel magnetic field coupling only to the spin, not to the orbital motion of electrons~\cite{RN1,RN6}, polarizes the graphene carriers, affecting the density of states~\cite{RN2} and thereby tuning the resistivity. The Coulomb interaction and disorder are two important aspects in graphene systems that may induce intriguing physics~\cite{PhysRevLett.106.236805,Tang2018,Hesselmanneaav6869,PhysRevB.77.115109}. In the limit of field strength $B=0$, they can introduce metallic behavior into the system even at the Dirac point~\cite{RN5}, and in the QHE, their interplay with the magnetic field determines the combined four-flavor degeneracy~\cite{NaturePhysics.8.7}, which can be thought of as a single SU(4) isospin because of the high energy scales characterizing cyclotron motion and Coulomb interactions~\cite{RN124,RN125}. Therefore, studies on the magnetic field and disorder in interacting Dirac fermion systems may not only help us deepen the understanding of the internal physical mechanism of the MIT but also help us design new progress in the application of 2D materials.

In this article, we studied the Hubbard model on a honeycomb lattice through the exact determinant quantum Monte Carlo (DQMC) method. Our data suggest that the Zeeman effect suppresses metallic behavior and bring about a transition from a metallic phase to an insulating one at a critical field strength, similar to what is seen in 2D holes in GaAs~\cite{PhysRevLett.84.5592}. This phenomenon occurs even with weak disorder and interaction. Different from the Mott or Anderson insulators (MI or AI), a sufficiently strong magnetic field opens a gap by affecting the electrons with different spins, and thereby inducing a band insulator. By making use of the density of states at the Fermi level in the limit that the temperature $T \to $ 0, we unambiguously determine the type of insulator~\cite{PhysRevLett.117.146601}, arising from the competition of interaction, disorder and magnetic field in the phase diagram [Fig.~\ref{Fig1}(a)] of the model, which further displays metallic behavior at sufficiently small magnitudes of these terms.

Besides, the impact of these three `knobs' on the transport properties is not isolated. Reducing the interaction and disorder makes the influence of magnetic field more pronounced. In contrast, switching on the magnetic field makes electrons to be more easily localized, corresponding to a decrease in the critical disorder or interaction strength for the corresponding metal-insulator transition. Their interplay is reflected in Figs.~\ref{Fig3} and~\ref{Fig4}. Interestingly, although entering a fully spin-polarized state often means a sudden change in transport properties, the critical field strength for the $B$-driven phase transition we investigate does not coincide with that of full spin polarization, being actually much smaller. This suggests that a weak magnetic field may promote a metal-insulator transition with lower interaction strengths and smaller disorder.

\section{Model and method}
\begin{figure}[t]
\flushleft
\begin{minipage}[c]{0.23\textwidth}
\centering
\includegraphics[width=5.0cm]{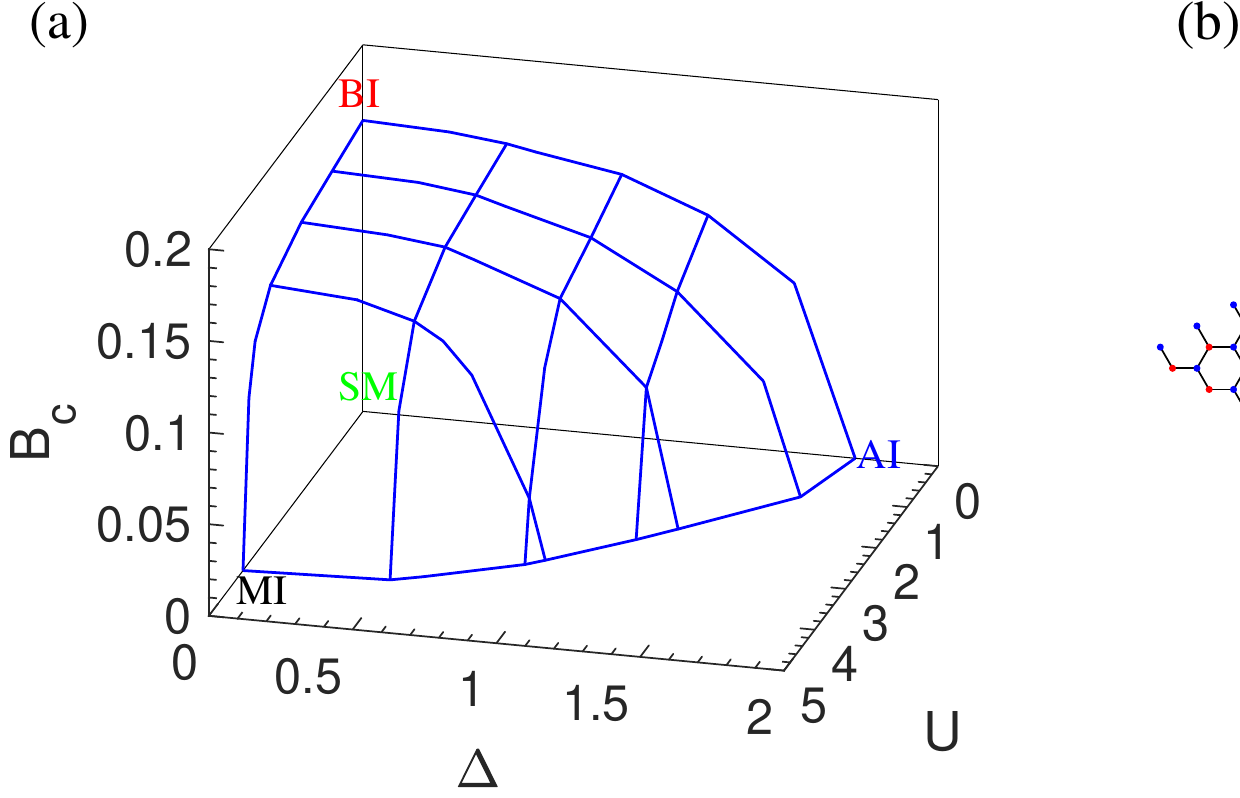}
\end{minipage}
\begin{minipage}[c]{0.2\textwidth}
\centering
\includegraphics[width=4.0cm]{Fig1b}
\end{minipage}
\caption{(a) Phase diagram of the disordered Hubbard model of a honeycomb lattice at half-filling with magnetic field $B$. $\Delta$ labels the disorder, $U$ the local Coulomb repulsion and $B_{c}$ the critical magnetic field. The points on the surface represent the MIT, and crossing the surface from below to above means that the metallic system enters the insulating phase driven by the interaction, disorder or magnetic field.
(b) The picture shows a $2\times 12^2$ honeycomb lattice with periodic boundary conditions, whose sublattices are labeled by red and blue colors. The applied magnetic field is parallel to the plane of the crystal lattice, which only affects the spin polarization of the electrons and does not affect the degeneracy of the orbital angular momentum.}
\label{Fig1}
\end{figure}

The Hamiltonian of the disordered Hubbard model on a honeycomb lattice on the presence of a magnetic field is defined as:
\begin{eqnarray}
\label{Hamiltonian}
\hat{H}&=&-\sum_{\langle{\bf ij}\rangle\sigma}t_{\bf ij}(\hat{c}_{{\bf i}\sigma}^\dagger\hat{c}_{{\bf j}\sigma}
+\hat{c}_{{\bf j}\sigma}^\dagger\hat{c}_{{\bf i}\sigma})+U\sum_{\bf j}(\hat{n}_{{\bf j}\uparrow}-\frac{1}{2})(\hat{n}_{{\bf j}\downarrow}-\frac{1}{2}) \nonumber\\
&&-\sum_{{\bf j}\sigma}(\mu-{\sigma}B)\hat{n}_{{\bf j}\sigma},
\end{eqnarray}
where $\hat{c}_{{\bf j}\sigma}^\dagger(\hat{c}_{{\bf j}\sigma})$ is the spin-$\sigma$ electron creation (annihilation) operator at site $\bf i$, and $\hat{n}_{{\bf i}\sigma}=\hat{c}_{{\bf i}\sigma}^\dagger\hat{c}_{{\bf i}\sigma}$ is the occupation number operator --- See Fig.~\ref{Fig1}(b) for the lattice schematics. Here, $t_{\bf ij}$ is the \textit{nearest-neighbor} (NN) hopping integral, $U > 0$ is the onsite Coulomb repulsion, $\mu$ is the chemical potential, and $B$ is the Zeeman magnetic field along the lattice plane (thus not generating orbital contributions~\cite{RN1}). Disorder is introduced through the hopping parameters $t_{\bf ij}$ taken from the probability distribution ${\cal P}(t_{\bf ij})$ = 1$/\Delta$ for $t_{\bf ij} \in [t - \Delta/2,t + \Delta/2]$ and zero otherwise. $\Delta$ describes the strength of disorder, and $t=1$ sets the energy scale in what follows. By choosing $\mu=0$, the system is half-filled, and particle-hole symmetry takes place~\cite{RN101}.

We adopt the DQMC method~\cite{RN102} to study the MIT in the model defined by Eq.~\eqref{Hamiltonian}, in which the Hamiltonian is mapped onto free fermions in 2D+1 dimensions coupled to space- and imaginary-time-dependent bosonic (Ising-like) fields. By using Monte Carlo sampling, we can carry out the integration over a relevant sample of field configurations, chosen up until statistical errors become negligible. The discretization mesh $\Delta\tau$ of the inverse temperature $\beta = 1/T$ should be small enough to ensure that the Trotter errors are less than those associated with the statistical sampling. This approach allows us to compute static and dynamic observables at a given temperature $T$. Due to the particle-hole symmetry even in the presence of the hopping-quenched disorder, the system avoids the infamous minus-sign problem, and the simulation can be performed at large enough $\beta$ as to obtain properties converging to the ground state ones~\cite{RN5,RN115}. We choose a honeycomb lattice with periodic boundary conditions, whose $L = 12$ geometry is shown in Fig.~\ref{Fig1} (b), and the total number of sites is $N =2\times$$L^2$. In the presence of disorder, we average over 20 disorder realizations~\cite{RN5,RN103a,RN103b,RN103c,RN103d} -- see Appendix~\ref{app:fse} for a system size comparison and Appendix~\ref{app:realz} for the impact of realization averaging.

The $T$-dependent dc conductivity is computed via a proxy of the momentum $\bf q$ and imaginary time $\tau$-dependent current-current correlation function (see Appendix~\ref{app:dc_formula}):
\begin{eqnarray}
\label{conductivity}
\sigma_{\rm dc}(T)=\frac{\beta^{2}}{\pi}\Lambda_{xx}\left({\bf q}=0,\tau=\frac{\beta}{2}\right).
\end{eqnarray}
Here, $\Lambda_{xx}({\bf q},\tau)$ = $\langle \widehat{j_{x}}({\bf q},\tau)\widehat{j_{x}}(-{\bf q},0)\rangle$, and $\widehat{j_{x}}({\bf q},\tau)$ is the current operator in the $x$ direction. This form, which avoids the analytic continuation of the QMC data, has been seen to provide satisfactory resuls for either disordered ~\cite{Trivedi1996,Scalettar1999,RN5} or clean systems~\cite{Mondaini2012,Lederer2017,Huang2019}.

By the same token, we define $N(0)$, the density of states at the Fermi level, as~\cite{RN104,Lederer2017}
\begin{eqnarray}
\label{N0}
N(0)\simeq \beta \times G({\bf r} = 0,\tau = \beta/2),
\end{eqnarray}
to differentiate the several physical mechanisms responsible for inducing the insulating phase, where $G$ is the imaginary-time dependent Green's function. We finally introduce the parameter $P=|n_{\downarrow}-n_{\uparrow}|/(n_{\downarrow}+n_{\uparrow})$ to study the spin polarization of electrons, where $n_{\downarrow}$ and $n_{\uparrow}$ are the averaged spin-resolved densities of the corresponding number operators in Eq.~\eqref{Hamiltonian}.

\begin{figure}[b]
\centering
\includegraphics[width=8cm]{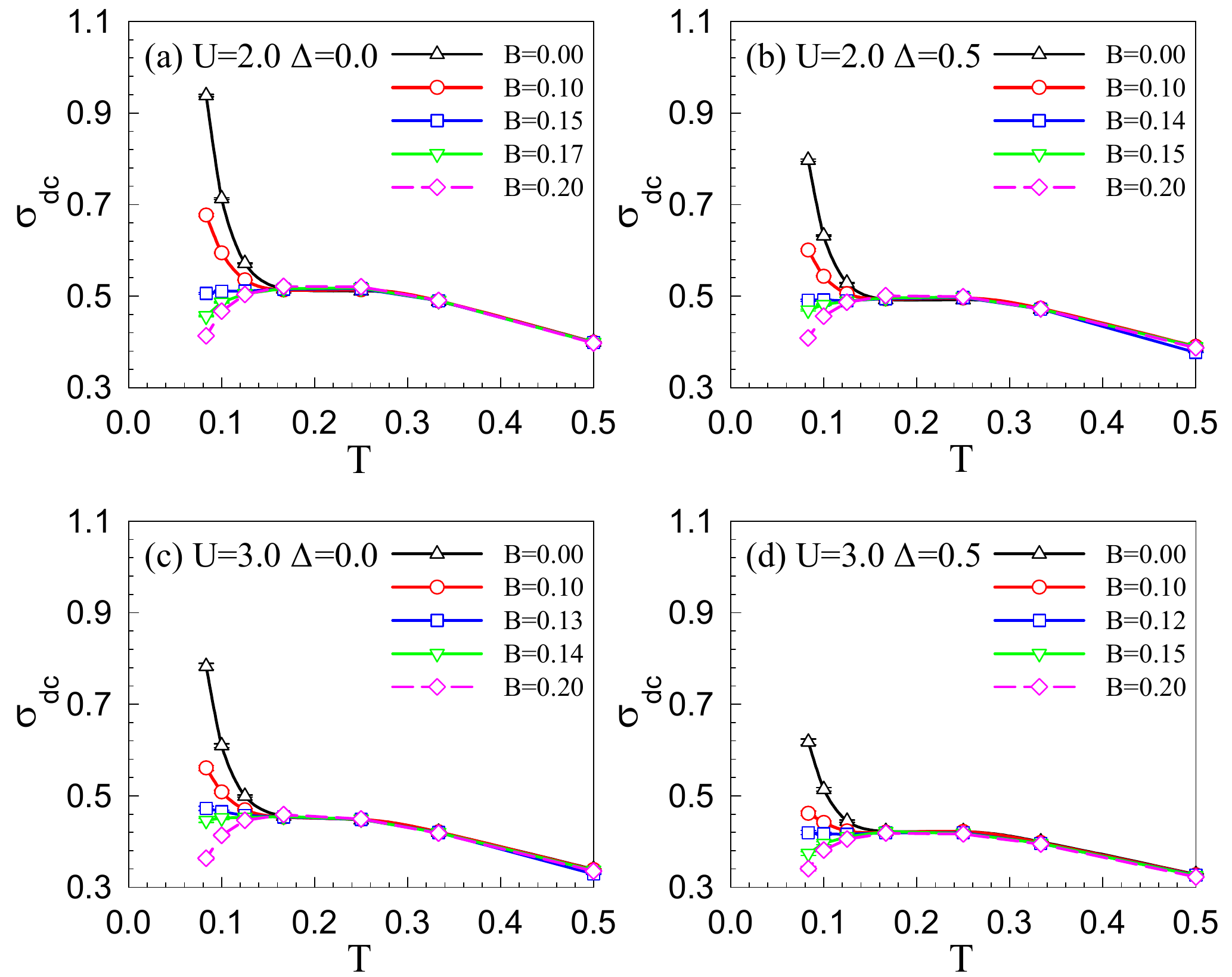}
\caption{Conductivity $\sigma_{\rm dc}$ as a function of temperature $T$ for various strengths of the Zeeman field, Coulomb interaction and disorder. The critical value of the magnetic field $B_{c}$, which indicates the occurrence of the transition from metallic to insulating behavior, is estimated to be $\simeq$ 0.15, 0.14, 0.13 and 0.12 at $U = 2.0$, $\Delta$ = 0.0 (a); $U = 2.0$, $\Delta = 0.5$ (b); $U = 3.0$, $\Delta = 0.0$ (c); and $U = 3.0$, $\Delta = 0.5$ (d), respectively.}
\label{Fig:dcUB}
\end{figure}

\section{Result and Discussion} We start by reporting the $\sigma_{\rm dc}(T)$ computed across several representative sets of $U$ and $\Delta$ and various fields $B$ in Fig.~\ref{Fig:dcUB}. While the conductivity decreases at lower temperatures in the (semi-)metallic phase with sufficient small interaction and disorder values, the effect of an increase in magnetic field is unequivocal: it induces a suppression of metallic behavior, displaying a downturn of $\sigma_{\rm dc}$ at small $T$'s. It confirms a $B$-driven MIT, verified in all panels in Fig.~\ref{Fig:dcUB} (exhibiting different combinations of $(U,\Delta)$), whose results could be potentially connected to what is observed in thin films of GaAs, also featuring a two-dimensional hexagonal lattice~\cite{PhysRevLett.84.5592}. A precise definition of the critical magnetic field $B_{c}(U,\Delta)$ that triggers the MIT is then obtained via d$\sigma_{\rm dc}/$d$T=0$ at low temperatures. On a similar fashion, we can also obtain the $\Delta_c$ and $U_c$ for the onset of the MIT, observing the signal of d$\sigma_{\rm dc}/$d$T$.

\begin{figure}[t]
\includegraphics[width=8.5cm]{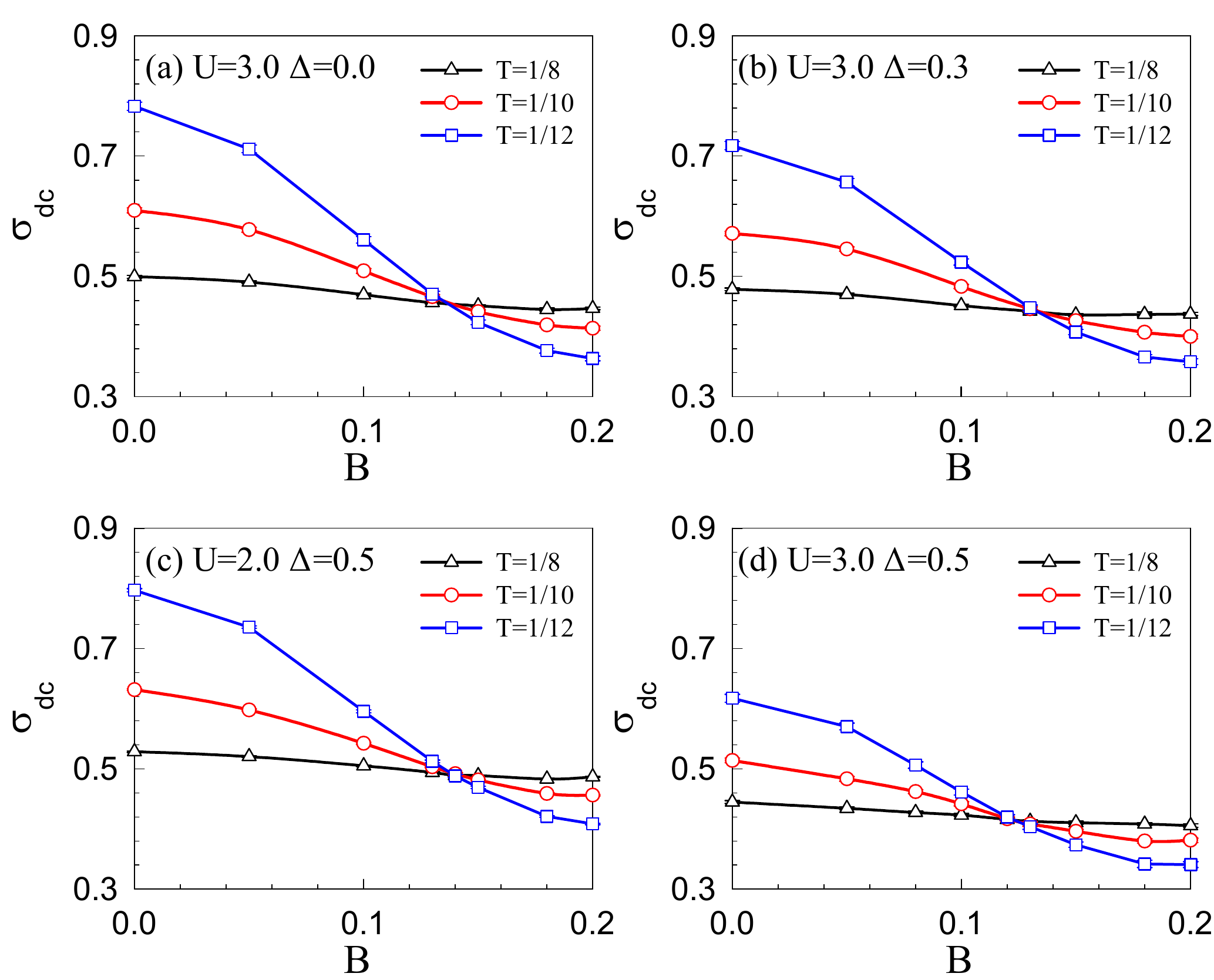}
\caption{\label{Fig3} dc conductivity as a function of magnetic field $B$ at various temperatures with (a) $U=3.0$, $\Delta=0.0$; (b) $U=3.0$, $\Delta=0.3$; (c) $U=2.0$, $\Delta=0.5$; and (d) $U=3.0$, $\Delta=0.5$. The effect of the magnetic field is more obvious with weaker interaction and disorder strength. With the change of $U$ and $\Delta$, the positions of the intersection points move quite obviously; $B_{c}$ in (a), (b), (c) and (d) is approximately 0.136, 0.132, 0.142 and 0.124 respectively.}
\end{figure}

\begin{figure}[b]
\includegraphics[width=9cm]{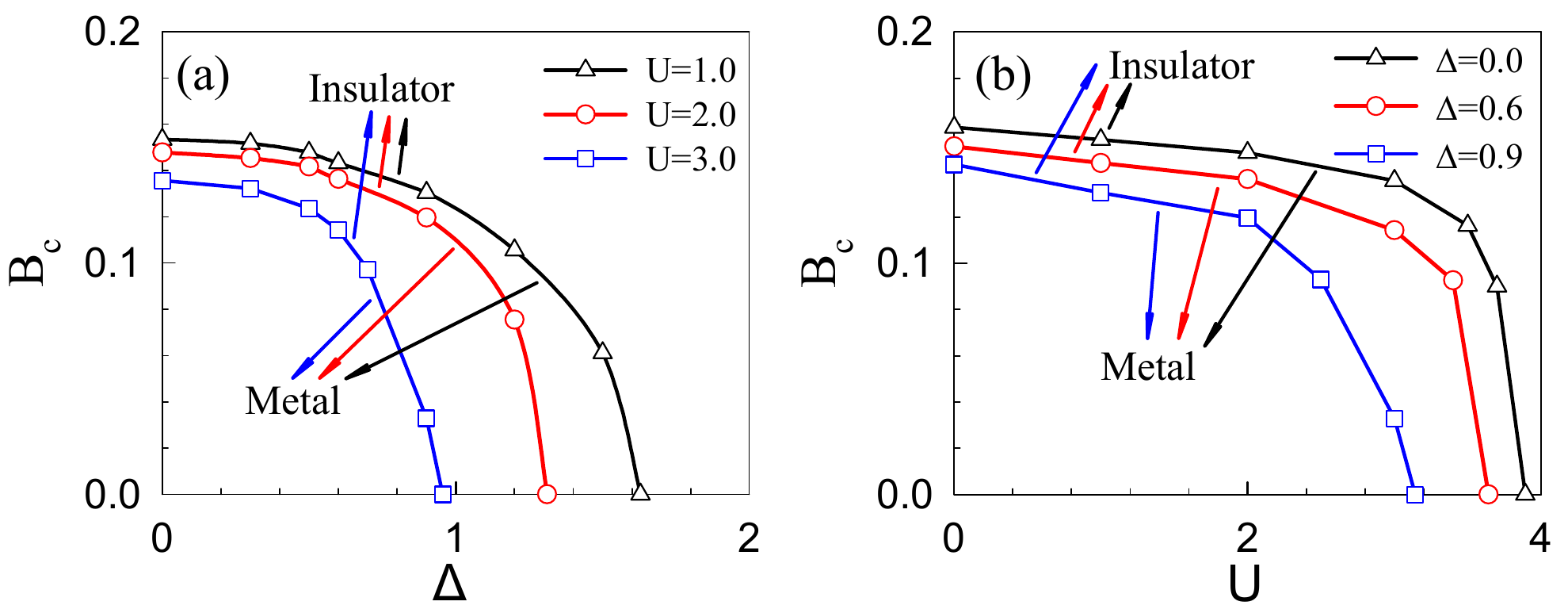}
\caption{\label{Fig4} Critical magnetic field strength $B_{c}$ (a) as a function of $\Delta$ at different $U$ and (b) as a function of $U$ at different $\Delta$. These cuts represent how the phase diagram in Fig.~\ref{Fig1}(a) is constructed. The intersection points of the curves with the horizontal axis indicate that (a) $\Delta$ is large enough to induce an Anderson insulator (at $U$ = 1.0, 2.0, and 3.0, $\Delta_{c}$ is approximately 1.63, 1.31, and 0.96) and (b) $U$ is large enough to induce a Mott insulator (at $\Delta$ = 0.0, 0.6, and 0.9, $U_{c}$ is approximately 3.89, 3.64, and 3.14).}
\end{figure}

A more evident display of the critical magnetic field is obtained in Fig.~\ref{Fig3}, where the dc conductivity at different temperatures $T$ is shown as a function of $B$ -- see Appendix~\ref{app:U_dis_transition} for a similar analysis with growing interaction strengths instead. In all cases, $\sigma_{\rm dc}$ monotonically decreases with increasing magnetic field, where the intersection of the curves defines the critical field $B_c$ for different $U$ and $\Delta$. We caution though that this determination of the $B$-driven MIT point is slightly more problematic when both $U$ and $\Delta$ are large, where often a region of intersections is observed for the different fixed temperature curves. Nonetheless, when increasing $\Delta$, as shown in Fig.~\ref{Fig3} (b) to Fig.~\ref{Fig3} (d), or increasing $U$, as shown in Fig.~\ref{Fig3} (c) to Fig.~\ref{Fig3} (d), a clear change on the position of the intersection point is observed, moving to smaller $B$'s. This suggests that $B_{c}$ is reduced either when the interaction or disorder are enhanced. These results also show that the magnetic field more strongly inhibits the metallic phase at lower temperatures, weaker interactions and weaker disorder. A similar phenomenon also occurs in a real material, hydrogenated graphene~\cite{PhysRevB.97.161402}.

Lastly, we compile in Fig.~\ref{Fig4} the results of $B_{c}$, showing that it decreases as disorder $\Delta$ or interaction strength $U$ are enhanced, representing the separation of the metallic and insulating phases. An important contrast is the sharp drop of the critical field when entering the $U$-driven Mott insulating phase [Fig.~\ref{Fig4}(b)], in comparison to a more gradual evolution of $B_c$ when impacted by the disorder $\Delta$ [Fig.~\ref{Fig4}(a)]. When summing all these results, we can compile the whole phase diagram as shown in Fig.~\ref{Fig1}(a), summarizing the interplay between the magnetic field, interaction and disorder.

\begin{figure}[t]
\includegraphics[width=8cm]{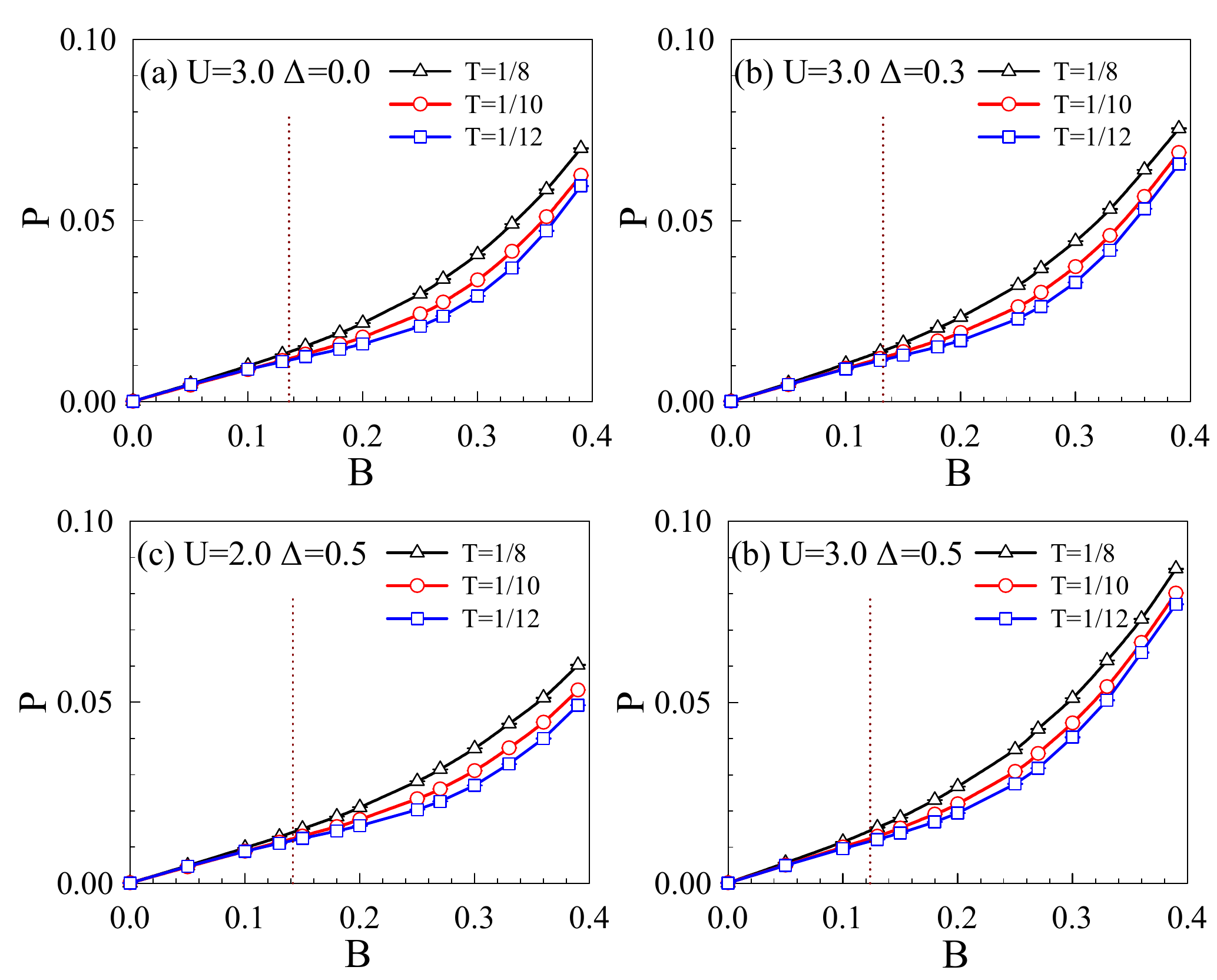}
\caption{\label{Fig5}Degree of spin polarization $P$ as a function of $B$ at low temperature. In panels (a) to (d), dashed lines depict the MIT, located approximately at 0.14, 0.13, 0.14, and 0.12, respectively. The values of $P$ at the MIT in these conditions are all smaller than 0.02, which are much lower than the value for full spin polarization.}
\end{figure}

It is worth noting that, in real materials as graphene, the estimated $U$ is relatively small~\cite{PhysRevLett.106.236805, PhysRevLett.111.036601} and metallic behavior, in opposition to a Mott phase, ensues. Our results suggest however that switching on a parallel magnetic field will decrease the critical strength for the MIT for both the interaction and disorder, providing a possibility of observing an interaction-driven phase transition. For example, for the MIT at $U = 3.14$ and $\Delta = 0.9$, applying a magnetic field $B \approx 0.33$ reduces the critical $U$ to 3.0. For the MIT at $U = 3.0$ and $\Delta$ = 0.96, this magnetic field reduces the critical $\Delta$ to 0.9. This phenomenon suggests that electron localization becomes more effective as the magnetic field increases, and we will now investigate the influence of spin polarization in these results.

In the limit that $B = 0$, $n_{\uparrow}$ and $n_{\downarrow}$ are equal, and by the interplay of interaction and hopping energy, electrons with opposite spins on NN sites can hop. If now introducing a magnetic field, the increased spin polarization $P$ favors localization in the system due to the onset of Pauli blockade, which prevents the now more likely same-spin electrons to conduct. In this regime, the transport is more influenced on the maximum of $n_{\uparrow}$ and $n_{\downarrow}$ rather than on the total electron density $n = n_{\uparrow} + n_{\downarrow}$~\cite{PhysRevB.52.5289,PhysRevB.97.161402}.

\begin{figure}[b]
\includegraphics[width=8.5cm]{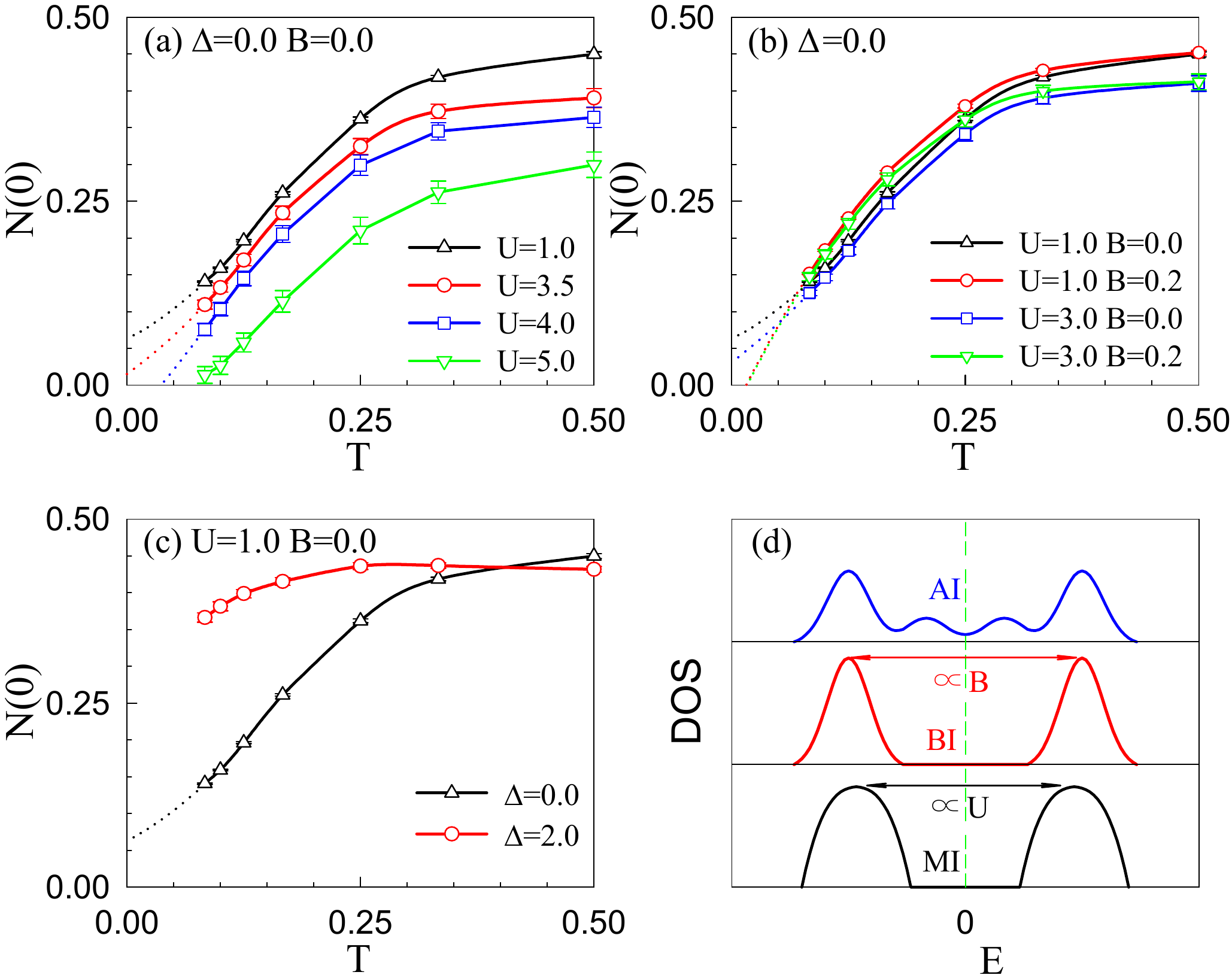}
\caption{\label{Fig6} Density of states at the Fermi energy $N(0)$, as a function of temperature $T$: (a) At various $U$. With $U$ increasing, $N(0)$ gradually decreases, and when $N(0)$ at $T \rightarrow 0$ tends to zero, the system becomes a Mott insulator. (b) At various $U$ and $B$. A sufficiently large $B$ can induce the system into a band insulating phase, whose $N(0)$ tends to zero at $T \rightarrow 0$. (c) At various $\Delta$. $N(0)$ of the Anderson insulator at low $T$ tends to a finite value. The MIT appears near $\Delta \approx$ 1.6. (d) Schematic diagram of the density of states for different insulator types. In panels (a), (b) and (c), we use data at low temperature to fit values of $N(0)$ near $T =$ 0. Since conditions at even lower temperatures are challenging, our polynomial fittings at $T \rightarrow$ (dashed lines) are to be interpreted on a qualitative level.}
\end{figure}

Figure~\ref{Fig5} shows the computed $P$ as a function of $B$ for several $(U,\Delta)$ combinations. The polarization degree of the system increases with both $U$ and $\Delta$, an opposite trend if compared to the dc conductivity. $P$ is always positively correlated to temperature, increases monotonically with the magnetic field, and, most importantly, \textit{does not} show special properties near $B_{c}$. Furthermore, $P$ at the MIT is much smaller than 1, indicating that the system is far from full-spin polarization. For instance, at $U$ = 3.0 and $\Delta = 0.0$, $B_{c}$ is approximately 0.14, which is much smaller than the field strength for full spin polarization [see Fig.~\ref{Fig5}(a)]. The corresponding value of $P$ is only $10^{-2}$, which suggests that the MIT does not coincide with the occurrence of a fully spin-polarized state~\cite{RN1,RN11,RN3}, and a small critical magnetic field strength is sufficient for the onset of the phase transition.

So far we have discussed how the different terms in~\eqref{Hamiltonian} drive a MIT, based on the analysis of $\sigma_{\rm dc}$. What this analysis misses is the differentiation, beyond a qualitative level, of the three different types of insulating phases one may reach. To better contrast those, we show in Fig.~\ref{Fig6} the density of states at the Fermi level $N(0)$ around the transitions driven by either $U$, $\Delta$ and $B$, discussing the different physical mechanisms through which they induce the corresponding insulating state [see Fig.~\ref{Fig6}(d) for schematics]. The interaction-induced Mott insulator, characterized by the opening of a Mott gap, results that $N(0)$ tends to 0 when $T \rightarrow$ 0~\cite{PhysRevLett.117.146601}, which is indeed observed near $U \approx$ 3.9 in Fig.~\ref{Fig6}(a). In turn, the disorder-driven Anderson insulator appears near $\Delta \approx$ 1.6 in Fig.~\ref{Fig6}(c), whose $N(0)$ is always finite at $T \rightarrow$ 0~\cite{PhysRevLett.117.146601}. Lastly, in panel (b), when $B$ is increased from 0.0 to 0.2 at $U = 1$, $\Delta = 0$ and $U = 3$, $\Delta = 0$, $N(0)$ at $T \rightarrow$ 0 changes from a finite value to 0. It indicates the formation of a band-insulating phase formed by the imbalance of the density with different spins.

\section{Summary}

Using DQMC simulations, we studied the metal-insulator transition of the disordered Hubbard model induced by a magnetic field on a honeycomb lattice. The parallel magnetic field, coupling to the electron spin, suppresses metallic behavior at low temperatures, and therefore induces the transition from a conducting to an insulating phase. We defined $B_{c}$ as the critical magnetic field at which the dc conductivity does not change in the low-temperature region. $B_{c}$ is overall much smaller than the strength required for full spin polarization, being further affected by the Coulomb repulsion and disorder: it reaches a maximum at very small $U$ and $\Delta$ and then displays an accelerated downward trend upon reaching close to the Anderson and Mott insulating phases. As a result, the conductance is largely influenced by the interplay of the three `knobs' we studied. For instance, the magnetic field has a more pronounced effect at small $U$ and $\Delta$, and in turn, the application of $B$ will effectively reduce the critical $U$ or $\Delta$ for the system to become a Mott insulator or an Anderson insulator. Right before the Mott insulating phase at $U=3.89$~\footnote{Note that this value is fairly close to the ones obtained directly at the ground state for much larger lattices~\cite{Sorella2012}, attesting the overall small finite size effects in the conductivity.} and $\Delta = 0.0$ [Fig.~\ref{Fig4}(b)], turning on $B$ has an initially insignificant effect, in which as $B$ increases considerably, the critical $U$ marginally decreases. Under a stronger magnetic field, however, the effect is greatly enhanced. A similar phenomenon also appears in the influence of $B$ on the critical $\Delta$.

In fact, an important differentiation can be drawn from the conductance with and without a magnetic field in the presence of $U$ and $\Delta$. As the application of $B$ polarizes graphene carriers, the carrier density is affected having direct consequences on the transport~\cite{RN5,RN2}. A partial Pauli blockade mechanism was interpreted as the basis of this positive magnetorresistance, but under a material's perspective, especially in light of recent schemes of disorder manipulation that have been currently advanced~\cite{NatureMaterials2019.18.6}, one may wonder if this interplay can be experimentally investigated in order to directly observe the MIT we disclose here.

\noindent
\underline{\it Acknowledgments} ---
We thank Richard T. Scalettar for many helpful discussions.
This work was supported by the NSFC (Nos. 11974049, 11774033, 11974039, 12088101 and 12050410263), the Beijing Natural Science Foundation (No. 1192011) and the NSAF-U1930402. The numerical simulations were performed at the HSCC of Beijing Normal University and on the Tianhe-2JK in the Beijing Computational Science Research Center.

\appendix

\setcounter{equation}{0}
\setcounter{figure}{0}
\renewcommand{\theequation}{A\arabic{equation}}
\renewcommand{\thefigure}{A\arabic{figure}}
\renewcommand{\thesubsection}{A\arabic{subsection}}

\section{Finite size effects}
\label{app:fse}
\begin{figure}[htbp]
\includegraphics[width=8.0cm]{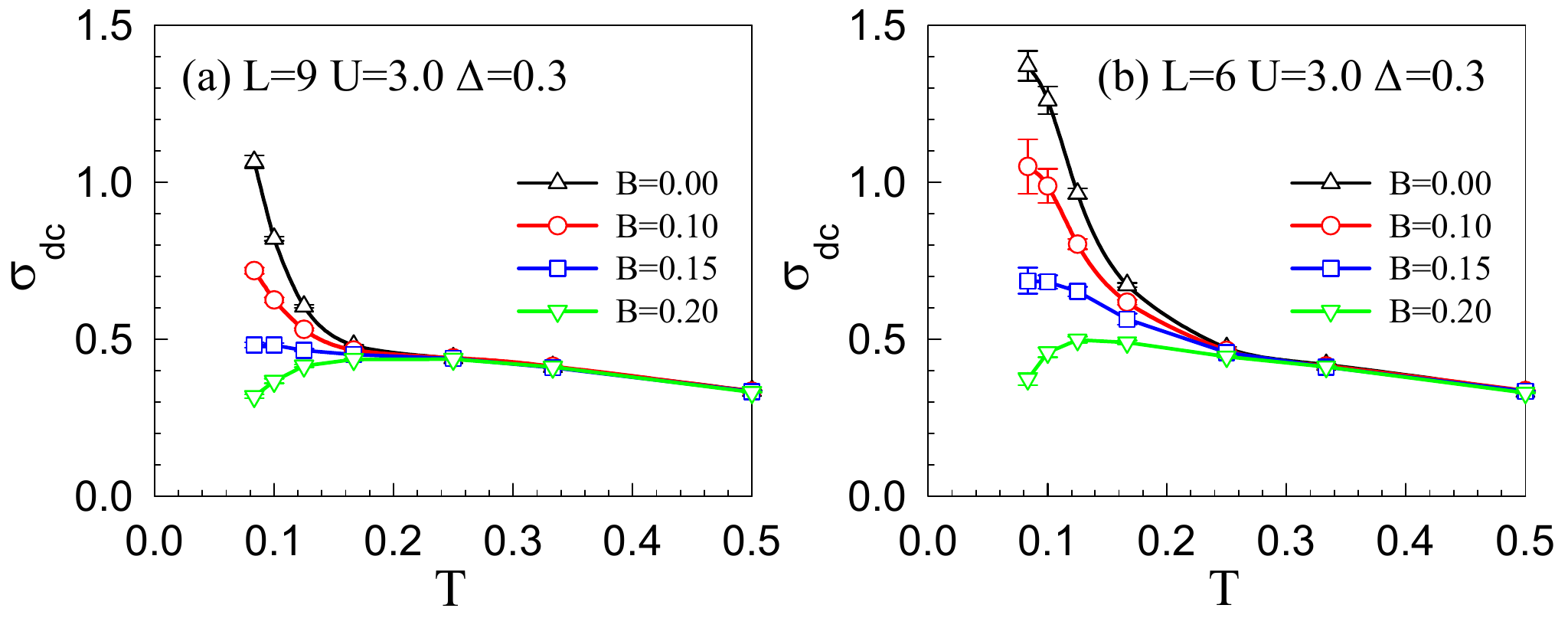}
\centering
\caption{The conductivity $\sigma_{\rm dc}$ is shown as a function of temperature $T$ at half-filling for various magnetic field strengths with lattice sizes (a) $L$ = 9 (b) $L$ = 6.
}
\label{SM1}
\end{figure}

To understand the influence of the system's finiteness on the physical results we have presented in the main text, we now check the fate of the conductivity $\sigma_{\rm dc}$ with various values of $L$. We start by reporting in Fig.~\ref{SM1} the conductivity $\sigma_{\rm dc}$ as a function of the temperature $T$ for the lattice sizes $L$ = 9 and 6.

\begin{figure}[b]
\includegraphics[width=8.0cm]{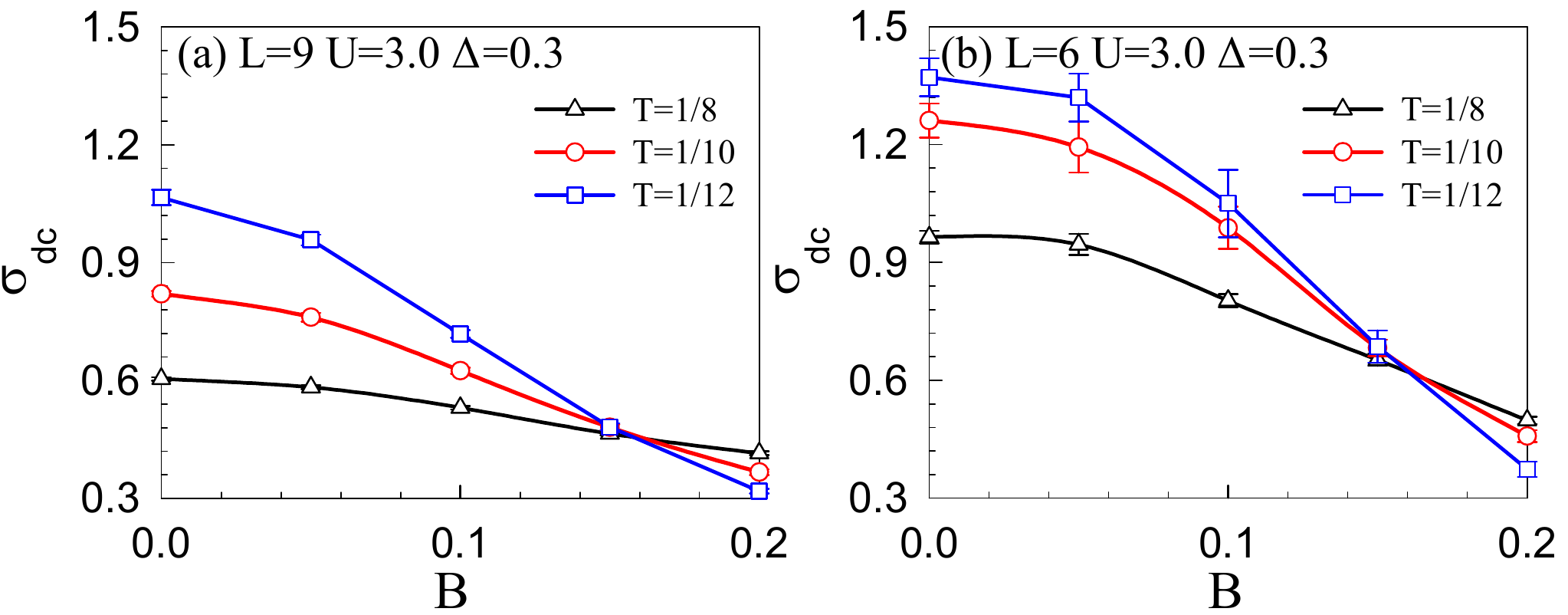}
\centering
\caption{The conductivity $\sigma_{\rm dc}$ is shown as a function of $B$ for various temperatures $T$ with (a) $L$ = 9 (b) $L$ =6.}
\label{SM2}
\end{figure}

While different lattice sizes yield different values for the conductivity ($\sigma_{\rm dc}$ grows as the system size decreases), the Zeeman field under all conditions still induces a band insulating phase at some critical value.

A more evident display is obtained in Fig.~\ref{SM2}, where $\sigma_{\rm dc}$ at different $T$'s is shown as a function of magnetic field $B$ for a set of $L$ values. In all cases, $\sigma_{\rm dc}$ monotonically decreases with increasing magnetic field. As before, the decrease in conductivity with growing lattice sizes is also observed, but more importantly, the critical value of the field associated with the metal-insulator transition (given by the intersection of the curves) is marginally dependent on the system size. As a result, finite size effects in the `surface' describing the phase diagram in Fig.~\ref{SM1}(a) in the main text are rather small.

\section{Interaction- or disorder-induced insulating phases}
\label{app:U_dis_transition}
\begin{figure}[t]
\includegraphics[width=8.5cm]{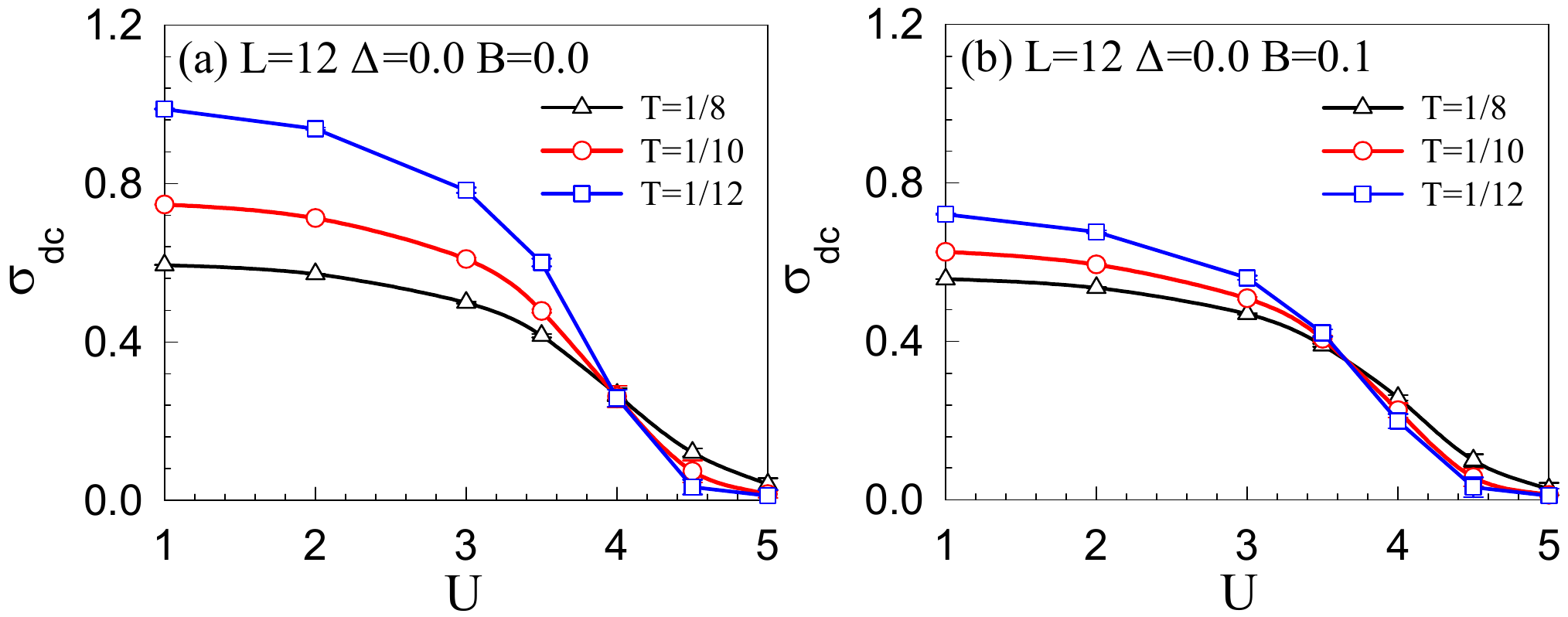}
\centering
\caption{The DC conductivity $\sigma_{\rm dc}$ is shown as a function of $U$ for various $T$ with (a) $B$ = 0.0 (b) $B$ = 0.1.}
\label{SM3}
\end{figure}

Our results suggest the impact of the three `knobs' on the transport properties is not isolated. The effect of the interaction $U$ and disorder $\Delta$ on magnetic field $B$ are shown in the main text, and here we provide data on how $B$ influences $U$, i.e., how the $U$-driven Mott transition is affected by the presence of a small magnetic field $B$. Figure~\ref{SM3} shows the $\sigma_{\rm dc}$ as a function of the $U$ in the clean case ($\Delta =$ 0) in a lattice with $L =$ 12. At low temperatures, the effect of an increase in $B$ is unequivocal: it not only suppresses the metallic behavior, but furthermore, it moves the position of the intersection point to smaller $U$'s (if one increases the magnetic field, this phenomenon will becomes even more pronounced). It means that the critical interaction $U_{c}$ is reduced when the magnetic field is included, and similar phenomenon also appears on the disorder-induced insulating phases. This suggests that localization becomes more effective as the magnetic field increases.

\section{The DC conductivity formula}
\label{app:dc_formula}

In this work, we use the low temperature behavior of DC conductivity $\sigma_{\rm dc}$ to distinguish metallic or insulating phases. We implemented the approach proposed in the Ref.\cite{RN103a},
which is based on the following argument. The fluctuation-dissipation theorem yields,
\begin{eqnarray}
\label{conductivity}
\Lambda_{xx}(\textbf{q},\tau)=\frac{1}{\pi}\int d\omega \frac{e^{-\omega \tau}}{1-e^{-\beta\omega}}\text{Im}\Lambda_{xx}(\textbf{q},\omega),
\end{eqnarray}
where $\Lambda_{xx}$ is the current-current correlation function along $x$-direction. While $\text{Im}\Lambda_{xx}(\textbf{q},\omega)$ could be computed by a numerical analytic continuation of $\Lambda_{xx}(\textbf{q},\tau)$ data obtained in DQMC, we instead here assume that $\text{Im}\Lambda_{xx}\sim\omega\sigma_{dc}$ below some energy scale $\omega < \omega^*$. Provided the temperature $T$ is sufficiently smaller than $\omega^*$, the above equation simplifies to
\begin{eqnarray}
\label{simpconduc}
\Lambda_{xx}\left(\textbf{q}=0,\tau=\frac{\beta}{2}\right)=\frac{\pi}{\beta^2}\sigma_{dc}
\end{eqnarray}
which is Eq.~\eqref{simpconduc} in the manuscript.

It has been noted that this approach may not be valid for a Fermi liquid\cite{RN103a}, when the characteristic energy scale is set by $\omega^* \sim N(0)T^2$, and the requirement $T<\omega^*$ will never be satisfied. However, in our system, the energy scale is set by the temperature-independent hopping-disorder strength $\omega^* \sim \Delta$, so that Eq.(\ref{simpconduc}) is valid at low temperatures.

\section{Concerning the number of disorder realizations} \label{app:realz}
\begin{figure}[htbp]
\includegraphics[scale=0.5]{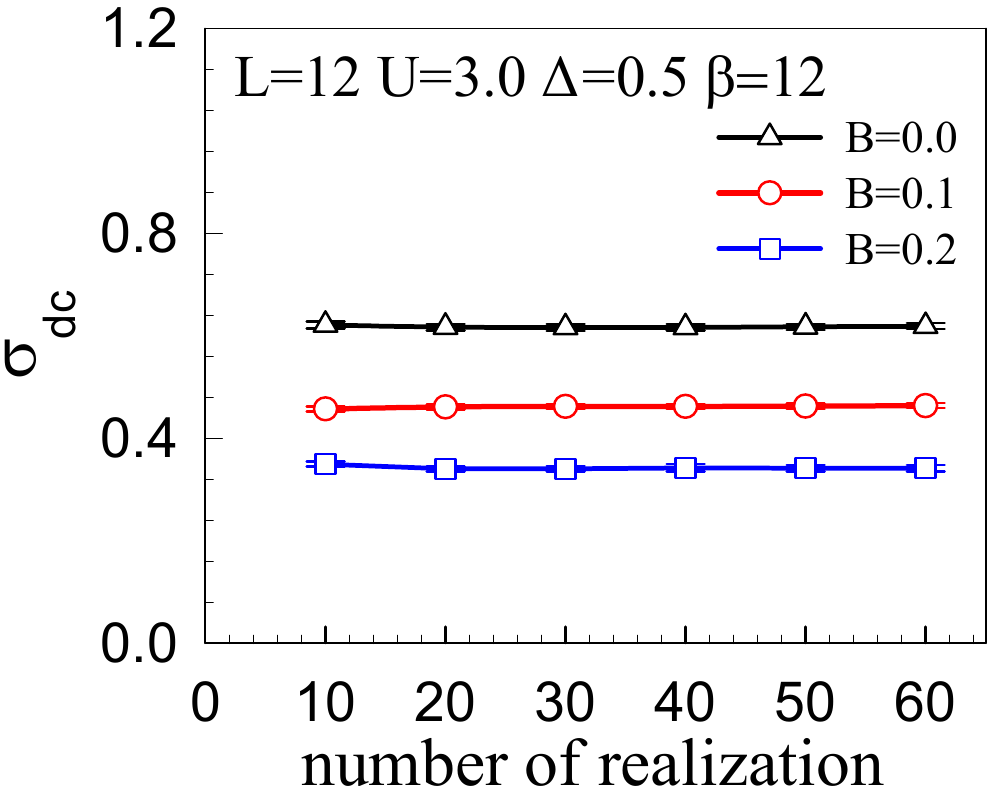}
\centering
\caption{DC conductivity computed on the $L =$ 12 lattice at $\beta=$12, $U=$3 and $\Delta=$0.5.
For a given $B$, the data obtained from an ensemble with growing disorder realizations are consistent within the statistical errors.
}
\label{SM4}
\end{figure}

In general, the required number of realizations in simulations with disorder must be determined empirically, which is a complex interplay between ``self-averaging'' on sufficiently large lattices, the disorder strength, and the location in the phase diagram. In Fig.~\ref{SM4}, we show the results of $\sigma_{\rm dc}$ averaged over different number of random disorder realizations. For any given magnetic field $B$, the averaged $\sigma_{\rm dc}$' s are already consistent for realization numbers larger than 10. It justifies the usage of 20 realizations which we have performed in the results in the main text. More precisely, our data suggest that there is considerable self-averaging on lattices with 2$L^{2} $= 288 sites.

\section{Canted antiferromagnetic phase}
In the absence of disorder ($\Delta=0$), Eq.\eqref{Hamiltonian} has been investigated at $T=0$ using unbiased projective QMC methods~\cite{PhysRevB.80.045412} in lattices with similar size as the ones we tackle here. They observe that in the semi-metallic phase, the in-plane magnetic field gives rise to a canted antiferromagnetic state, that is, one that displays a staggered magnetization perpendicular to the applied field for arbitrarily small values of the interactions.

\begin{figure}[t]
\vspace{0.5cm}
\includegraphics[width=0.99\columnwidth]{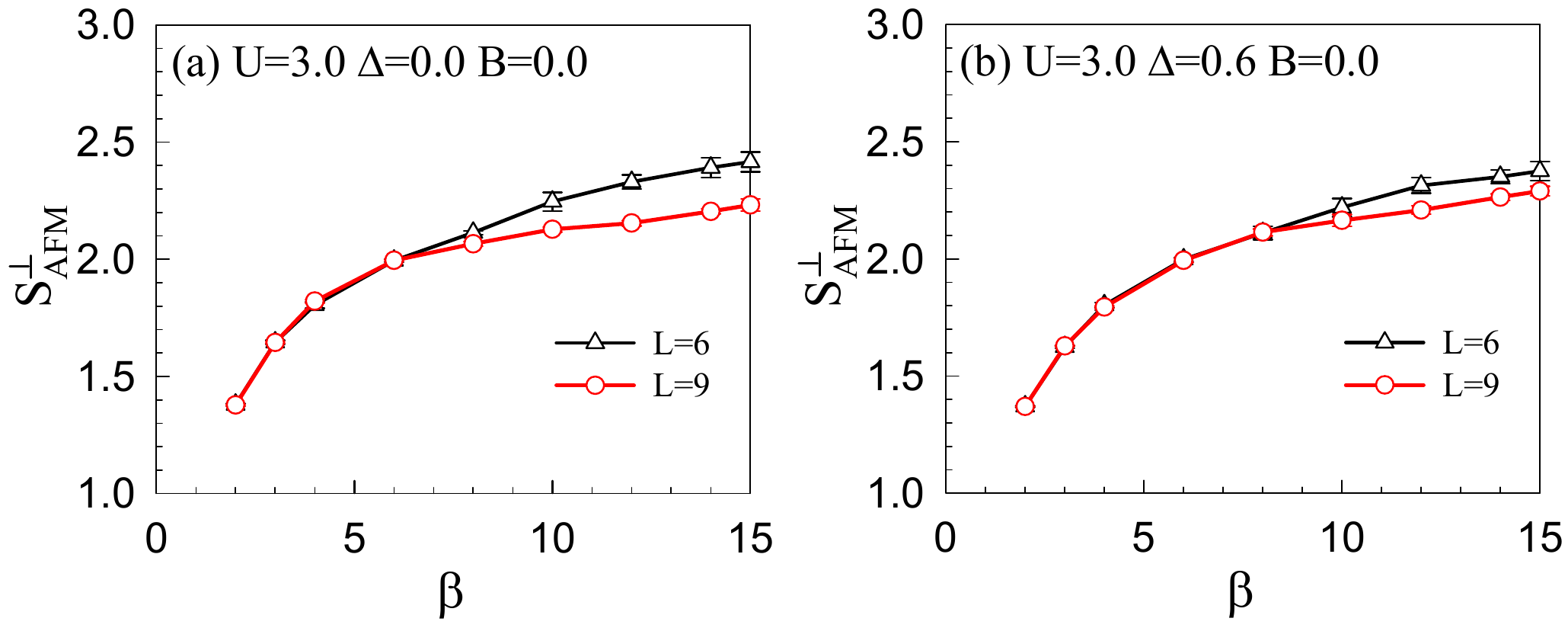}
\centering
\caption{Dependence of the transverse antiferromagnetic structure factor with the inverse temperature $\beta$ in the absence of field, for $\Delta=0$ (a) and $\Delta=0.6$ (b), and different lattice sizes. At $\beta=12$,  $S_{\rm AFM}^{\perp}$ is already close to the asymptotic value, denoting the results are close to the ones at $T=0$.}
\label{Fig_Appendix_S_beta}
\end{figure}

To characterize such phase, we compute the staggered transverse antiferromagnetic structure factor as
\begin{equation}
 S_{\rm AFM}^{\perp} = \frac{1}{N}\sum_{i,j}(-1)^{(i+j)} \left(S_i^{x}S_j^{x} + S_i^{y}S_j^{y}\right),
\end{equation}
where the phase factor is $+1$($-1$) for sites $i$,$j$ belonging to the same (different) sublattices of the honeycomb structure.
To test that at $\beta=12$ we are already assessing physics close to the ground-state, we show in Fig.~\ref{Fig_Appendix_S_beta} the dependence of $S_{\rm AFM}^{\perp}$ with the inverse temperature: saturation is readily observed for values $\beta\gtrsim12$.

\begin{figure}[t]
\includegraphics[width=0.99\columnwidth]{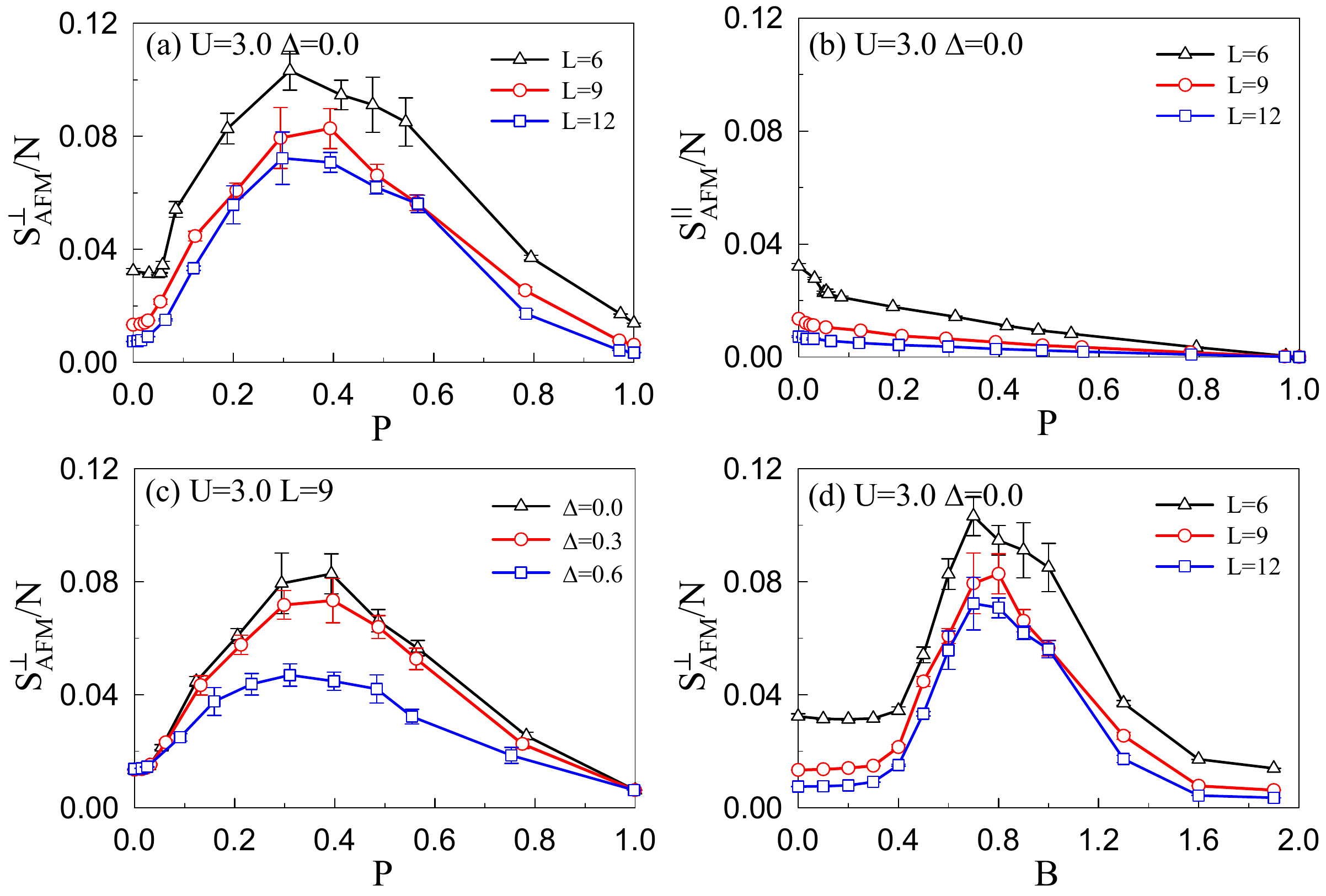}
\centering
\caption{(a) Dependence of the transverse antiferromagnetic structure factor on the polarization $P$ in the clean case; (d) the same but with a dependence on the $B$ field. (b) Similar results but disclosing the longitudinal antiferromagnetic structure factor at $\Delta=0$. (c) Influence of disorder on the transverse antiferromagnetic structure factor; at a given system size, a growing $\Delta$ diminishes the induced canted antiferromagnetic order. All data is taken at $\beta=12$ and interaction $U=3$.}
\label{Fig_Appendix_S_P}
\end{figure}

Fixing then at temperature $T=1/12$, we now focus on the dependence with the magnetic field, and the corresponding polarization $P$ in Fig.~\ref{Fig_Appendix_S_P}. In the clean case, the results of the transverse antiferromagnetic structure factor are remarkably similar to the ones from Ref.~\cite{PhysRevB.80.045412} when cast in terms of the polarization $P$ [Fig.~\ref{Fig_Appendix_S_P}(a)]. A contrast though is in order: projective QMC methods are canonical simulations and $P$ is an input of the calculation; here, however, a typical grand-canonical simulation, $P$ is an outcome that depends on the magnetic field used (and the remaining Hamiltonian's parameters). If converting these same results to the magnetic field strength, we notice that the regime where $S_{\rm AFM}^{\perp}$
quickly increases when tuning $B$ is much beyond the one that gives rise to the insulating transition, i.e., $S_{\rm AFM}^\perp$ only grows when the system is already in the insulating phase as quantified by the dc conductivity. In particular, for $U=3$ and $\Delta=0$, the critical field that drives the metal-insulating transition is $B_c=0.136$ [see Fig.~\ref{Fig3}(a)]. In fact, for this interaction magnitudes, increasing the disorder $\Delta$ dampens this effect, and the canted antiferromagnetic state becomes less prominent even at large field magnitudes/polarizations [see Fig.~\ref{Fig3}(c)].

Lastly, the longitudinal structure factor, $S_{\rm AFM}^{\parallel} \equiv (1/N)\sum_{i,j}(-1)^{(i+j)} S_i^{z}S_j^{z}$, is much smaller and quickly vanishes with growing lattice sizes below the critical interaction strength [see Fig.~\ref{Fig_Appendix_S_P}(b)].

\begin{figure}[t]
\includegraphics[width=0.99\columnwidth]{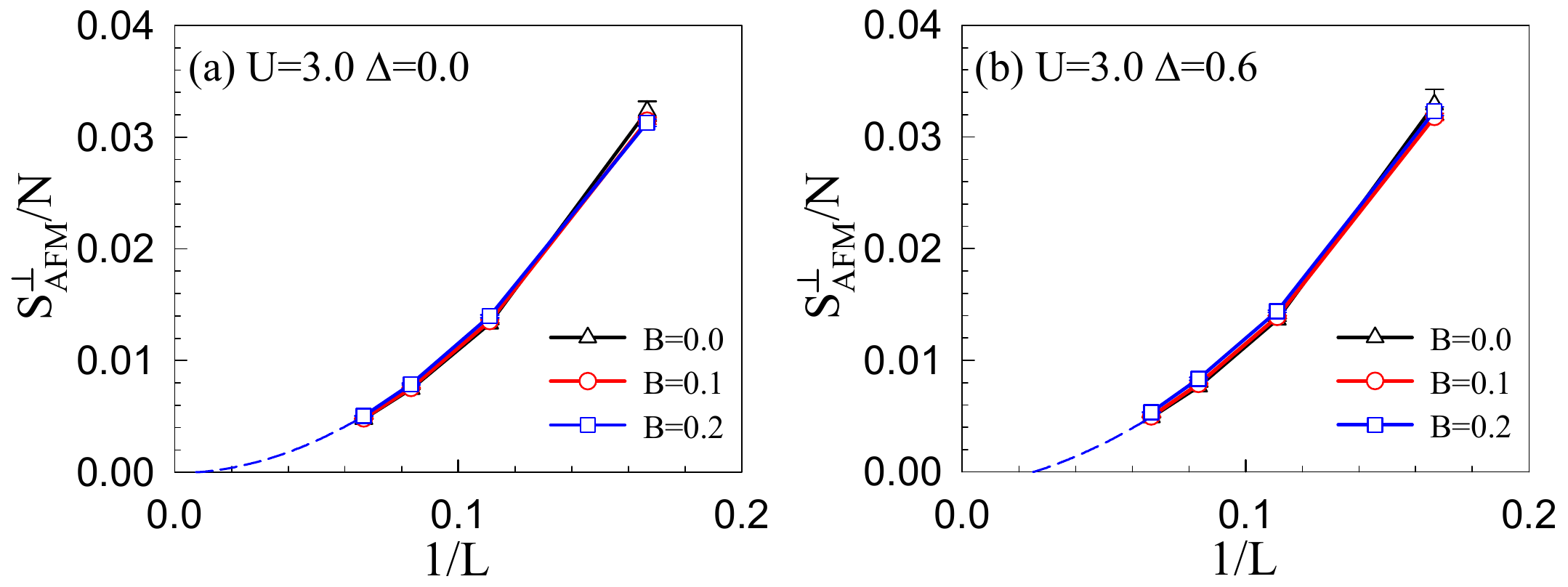}
\centering
\caption{(a) The finite size extrapolation of the normalized transverse structure factor in the clean case $\Delta=0$ (a) and the disordered one $\Delta=0.6$ (b), within a range of magnetic field strengths compatible with the phase diagram in Fig.~\ref{Fig1}(a). Dashed lines are polynomial fits to the points at $B=0.2$.}
\label{Fig_Appendix_Sxx_L}
\end{figure}

Now to verify whether the canted antiferromagnetic magnetic state survives when approaching the thermodynamic limit, we promote in Fig.~\ref{Fig_Appendix_Sxx_L} a finite-size analysis of the normalized $S_{\rm AFM}^{\perp}$, for values of the field concerned in the original phase diagram, Fig.~\ref{Fig1}, i.e., $B\lesssim0.2$. By scaling it with the inverse linear size, we notice that the canted antiferromagnetism does not appear in $L\to\infty$ limit in this regime of small fields, that is, its corresponding structure factor is not extensive, resulting in only short-range ordering. Further investigations would be necessary to study it at larger fields, and to test whether this state can overcome the disorder effects when approaching the thermodynamic limit.

\bibliography{Reference}

\begin{thebibliography}{57}%
\makeatletter
\providecommand \@ifxundefined [1]{%
 \@ifx{#1\undefined}
}%
\providecommand \@ifnum [1]{%
 \ifnum #1\expandafter \@firstoftwo
 \else \expandafter \@secondoftwo
 \fi
}%
\providecommand \@ifx [1]{%
 \ifx #1\expandafter \@firstoftwo
 \else \expandafter \@secondoftwo
 \fi
}%
\providecommand \natexlab [1]{#1}%
\providecommand \enquote  [1]{``#1''}%
\providecommand \bibnamefont  [1]{#1}%
\providecommand \bibfnamefont [1]{#1}%
\providecommand \citenamefont [1]{#1}%
\providecommand \href@noop [0]{\@secondoftwo}%
\providecommand \href [0]{\begingroup \@sanitize@url \@href}%
\providecommand \@href[1]{\@@startlink{#1}\@@href}%
\providecommand \@@href[1]{\endgroup#1\@@endlink}%
\providecommand \@sanitize@url [0]{\catcode `\\12\catcode `\$12\catcode
  `\&12\catcode `\#12\catcode `\^12\catcode `\_12\catcode `\%12\relax}%
\providecommand \@@startlink[1]{}%
\providecommand \@@endlink[0]{}%
\providecommand \url  [0]{\begingroup\@sanitize@url \@url }%
\providecommand \@url [1]{\endgroup\@href {#1}{\urlprefix }}%
\providecommand \urlprefix  [0]{URL }%
\providecommand \Eprint [0]{\href }%
\providecommand \doibase [0]{http://dx.doi.org/}%
\providecommand \selectlanguage [0]{\@gobble}%
\providecommand \bibinfo  [0]{\@secondoftwo}%
\providecommand \bibfield  [0]{\@secondoftwo}%
\providecommand \translation [1]{[#1]}%
\providecommand \BibitemOpen [0]{}%
\providecommand \bibitemStop [0]{}%
\providecommand \bibitemNoStop [0]{.\EOS\space}%
\providecommand \EOS [0]{\spacefactor3000\relax}%
\providecommand \BibitemShut  [1]{\csname bibitem#1\endcsname}%
\let\auto@bib@innerbib\@empty
\bibitem [{\citenamefont {Slater}(1951)}]{RN106}%
  \BibitemOpen
  \bibfield  {author} {\bibinfo {author} {\bibfnamefont {J.~C.}\ \bibnamefont
  {Slater}},\ }\href {\doibase 10.1103/PhysRev.82.538} {\bibfield  {journal}
  {\bibinfo  {journal} {Phys. Rev.}\ }\textbf {\bibinfo {volume} {82}},\
  \bibinfo {pages} {538} (\bibinfo {year} {1951})}\BibitemShut {NoStop}%
\bibitem [{\citenamefont {Crutcher}\ and\ \citenamefont
  {Kemball}(2019)}]{10.3389/fspas.2019.00066}%
  \BibitemOpen
  \bibfield  {author} {\bibinfo {author} {\bibfnamefont {R.~M.}\ \bibnamefont
  {Crutcher}}\ and\ \bibinfo {author} {\bibfnamefont {A.~J.}\ \bibnamefont
  {Kemball}},\ }\href {\doibase 10.3389/fspas.2019.00066} {\bibfield  {journal}
  {\bibinfo  {journal} {Front. Astron. Space Sci.}\ }\textbf {\bibinfo {volume}
  {6}},\ \bibinfo {pages} {66} (\bibinfo {year} {2019})}\BibitemShut {NoStop}%
\bibitem [{\citenamefont {Abanin}\ \emph {et~al.}(2011)\citenamefont {Abanin},
  \citenamefont {Gorbachev}, \citenamefont {Novoselov}, \citenamefont {Geim},\
  and\ \citenamefont {Levitov}}]{PhysRevLett.107.096601}%
  \BibitemOpen
  \bibfield  {author} {\bibinfo {author} {\bibfnamefont {D.~A.}\ \bibnamefont
  {Abanin}}, \bibinfo {author} {\bibfnamefont {R.~V.}\ \bibnamefont
  {Gorbachev}}, \bibinfo {author} {\bibfnamefont {K.~S.}\ \bibnamefont
  {Novoselov}}, \bibinfo {author} {\bibfnamefont {A.~K.}\ \bibnamefont {Geim}},
  \ and\ \bibinfo {author} {\bibfnamefont {L.~S.}\ \bibnamefont {Levitov}},\
  }\href {\doibase 10.1103/PhysRevLett.107.096601} {\bibfield  {journal}
  {\bibinfo  {journal} {Phys. Rev. Lett.}\ }\textbf {\bibinfo {volume} {107}},\
  \bibinfo {pages} {096601} (\bibinfo {year} {2011})}\BibitemShut {NoStop}%
\bibitem [{\citenamefont {Sun}\ \emph {et~al.}(2020{\natexlab{a}})\citenamefont
  {Sun}, \citenamefont {Song}, \citenamefont {Weng},\ and\ \citenamefont
  {Dai}}]{PhysRevB.101.125118}%
  \BibitemOpen
  \bibfield  {author} {\bibinfo {author} {\bibfnamefont {S.}~\bibnamefont
  {Sun}}, \bibinfo {author} {\bibfnamefont {Z.}~\bibnamefont {Song}}, \bibinfo
  {author} {\bibfnamefont {H.}~\bibnamefont {Weng}}, \ and\ \bibinfo {author}
  {\bibfnamefont {X.}~\bibnamefont {Dai}},\ }\href {\doibase
  10.1103/PhysRevB.101.125118} {\bibfield  {journal} {\bibinfo  {journal}
  {Phys. Rev. B}\ }\textbf {\bibinfo {volume} {101}},\ \bibinfo {pages}
  {125118} (\bibinfo {year} {2020}{\natexlab{a}})}\BibitemShut {NoStop}%
\bibitem [{\citenamefont {Dubi}\ \emph {et~al.}(2007)\citenamefont {Dubi},
  \citenamefont {Meir},\ and\ \citenamefont {Avishai}}]{Nature2007.449.7164}%
  \BibitemOpen
  \bibfield  {author} {\bibinfo {author} {\bibfnamefont {Y.}~\bibnamefont
  {Dubi}}, \bibinfo {author} {\bibfnamefont {Y.}~\bibnamefont {Meir}}, \ and\
  \bibinfo {author} {\bibfnamefont {Y.}~\bibnamefont {Avishai}},\ }\href
  {\doibase 10.1038/nature06180} {\bibfield  {journal} {\bibinfo  {journal}
  {Nature}\ }\textbf {\bibinfo {volume} {449}},\ \bibinfo {pages} {876}
  (\bibinfo {year} {2007})}\BibitemShut {NoStop}%
\bibitem [{\citenamefont {Sun}\ \emph {et~al.}(2020{\natexlab{b}})\citenamefont
  {Sun}, \citenamefont {Cao}, \citenamefont {Cui}, \citenamefont {Zhu},
  \citenamefont {Ma}, \citenamefont {Wang}, \citenamefont {Zhuo}, \citenamefont
  {Cheng}, \citenamefont {Wang}, \citenamefont {Wan},\ and\ \citenamefont
  {Chen}}]{Nature2020.5.1}%
  \BibitemOpen
  \bibfield  {author} {\bibinfo {author} {\bibfnamefont {Z.}~\bibnamefont
  {Sun}}, \bibinfo {author} {\bibfnamefont {Z.}~\bibnamefont {Cao}}, \bibinfo
  {author} {\bibfnamefont {J.}~\bibnamefont {Cui}}, \bibinfo {author}
  {\bibfnamefont {C.}~\bibnamefont {Zhu}}, \bibinfo {author} {\bibfnamefont
  {D.}~\bibnamefont {Ma}}, \bibinfo {author} {\bibfnamefont {H.}~\bibnamefont
  {Wang}}, \bibinfo {author} {\bibfnamefont {W.}~\bibnamefont {Zhuo}}, \bibinfo
  {author} {\bibfnamefont {Z.}~\bibnamefont {Cheng}}, \bibinfo {author}
  {\bibfnamefont {Z.}~\bibnamefont {Wang}}, \bibinfo {author} {\bibfnamefont
  {X.}~\bibnamefont {Wan}}, \ and\ \bibinfo {author} {\bibfnamefont
  {X.}~\bibnamefont {Chen}},\ }\href {\doibase 10.1038/s41535-020-0239-z}
  {\bibfield  {journal} {\bibinfo  {journal} {Nature}\ }\textbf {\bibinfo
  {volume} {5}},\ \bibinfo {pages} {36} (\bibinfo {year}
  {2020}{\natexlab{b}})}\BibitemShut {NoStop}%
\bibitem [{\citenamefont {Milat}\ \emph {et~al.}(2004)\citenamefont {Milat},
  \citenamefont {Assaad},\ and\ \citenamefont {Sigrist}}]{Milat2004}%
  \BibitemOpen
  \bibfield  {author} {\bibinfo {author} {\bibfnamefont {I.}~\bibnamefont
  {Milat}}, \bibinfo {author} {\bibfnamefont {F.}~\bibnamefont {Assaad}}, \
  and\ \bibinfo {author} {\bibfnamefont {M.}~\bibnamefont {Sigrist}},\ }\href
  {\doibase 10.1140/epjb/e2004-00154-5} {\bibfield  {journal} {\bibinfo
  {journal} {EPJ B}\ }\textbf {\bibinfo {volume} {38}},\ \bibinfo {pages} {571}
  (\bibinfo {year} {2004})}\BibitemShut {NoStop}%
\bibitem [{\citenamefont {Bercx}\ \emph {et~al.}(2009)\citenamefont {Bercx},
  \citenamefont {Lang},\ and\ \citenamefont {Assaad}}]{PhysRevB.80.045412}%
  \BibitemOpen
  \bibfield  {author} {\bibinfo {author} {\bibfnamefont {M.}~\bibnamefont
  {Bercx}}, \bibinfo {author} {\bibfnamefont {T.~C.}\ \bibnamefont {Lang}}, \
  and\ \bibinfo {author} {\bibfnamefont {F.~F.}\ \bibnamefont {Assaad}},\
  }\href {\doibase 10.1103/PhysRevB.80.045412} {\bibfield  {journal} {\bibinfo
  {journal} {Phys. Rev. B}\ }\textbf {\bibinfo {volume} {80}},\ \bibinfo
  {pages} {045412} (\bibinfo {year} {2009})}\BibitemShut {NoStop}%
\bibitem [{\citenamefont {Matsuda}\ \emph {et~al.}(2020)\citenamefont
  {Matsuda}, \citenamefont {Nakamura}, \citenamefont {Ikeda}, \citenamefont
  {Takeyama}, \citenamefont {Suga}, \citenamefont {Nakahara},\ and\
  \citenamefont {Muraoka}}]{Nature2020.11.1}%
  \BibitemOpen
  \bibfield  {author} {\bibinfo {author} {\bibfnamefont {Y.~H.}\ \bibnamefont
  {Matsuda}}, \bibinfo {author} {\bibfnamefont {D.}~\bibnamefont {Nakamura}},
  \bibinfo {author} {\bibfnamefont {A.}~\bibnamefont {Ikeda}}, \bibinfo
  {author} {\bibfnamefont {S.}~\bibnamefont {Takeyama}}, \bibinfo {author}
  {\bibfnamefont {Y.}~\bibnamefont {Suga}}, \bibinfo {author} {\bibfnamefont
  {H.}~\bibnamefont {Nakahara}}, \ and\ \bibinfo {author} {\bibfnamefont
  {Y.}~\bibnamefont {Muraoka}},\ }\href {\doibase 10.1038/s41467-020-17416-w}
  {\bibfield  {journal} {\bibinfo  {journal} {Nature}\ }\textbf {\bibinfo
  {volume} {11}},\ \bibinfo {pages} {3591} (\bibinfo {year}
  {2020})}\BibitemShut {NoStop}%
\bibitem [{\citenamefont {Imada}\ \emph {et~al.}(1998)\citenamefont {Imada},
  \citenamefont {Fujimori},\ and\ \citenamefont {Tokura}}]{RevModPhys.70.1039}%
  \BibitemOpen
  \bibfield  {author} {\bibinfo {author} {\bibfnamefont {M.}~\bibnamefont
  {Imada}}, \bibinfo {author} {\bibfnamefont {A.}~\bibnamefont {Fujimori}}, \
  and\ \bibinfo {author} {\bibfnamefont {Y.}~\bibnamefont {Tokura}},\ }\href
  {\doibase 10.1103/RevModPhys.70.1039} {\bibfield  {journal} {\bibinfo
  {journal} {Rev. Mod. Phys.}\ }\textbf {\bibinfo {volume} {70}},\ \bibinfo
  {pages} {1039} (\bibinfo {year} {1998})}\BibitemShut {NoStop}%
\bibitem [{\citenamefont {Melnikov}\ \emph {et~al.}(2020)\citenamefont
  {Melnikov}, \citenamefont {Shashkin}, \citenamefont {Dolgopolov},
  \citenamefont {Huang}, \citenamefont {Liu}, \citenamefont {Zhu},\ and\
  \citenamefont {Kravchenko}}]{PhysRevB.101.045302}%
  \BibitemOpen
  \bibfield  {author} {\bibinfo {author} {\bibfnamefont {M.~Y.}\ \bibnamefont
  {Melnikov}}, \bibinfo {author} {\bibfnamefont {A.~A.}\ \bibnamefont
  {Shashkin}}, \bibinfo {author} {\bibfnamefont {V.~T.}\ \bibnamefont
  {Dolgopolov}}, \bibinfo {author} {\bibfnamefont {S.-H.}\ \bibnamefont
  {Huang}}, \bibinfo {author} {\bibfnamefont {C.~W.}\ \bibnamefont {Liu}},
  \bibinfo {author} {\bibfnamefont {A.~Y.~X.}\ \bibnamefont {Zhu}}, \ and\
  \bibinfo {author} {\bibfnamefont {S.~V.}\ \bibnamefont {Kravchenko}},\ }\href
  {\doibase 10.1103/PhysRevB.101.045302} {\bibfield  {journal} {\bibinfo
  {journal} {Phys. Rev. B}\ }\textbf {\bibinfo {volume} {101}},\ \bibinfo
  {pages} {045302} (\bibinfo {year} {2020})}\BibitemShut {NoStop}%
\bibitem [{\citenamefont {Li}\ \emph {et~al.}(2019)\citenamefont {Li},
  \citenamefont {Zhang}, \citenamefont {Ghaemi},\ and\ \citenamefont
  {Sarachik}}]{PhysRevB.99.155302}%
  \BibitemOpen
  \bibfield  {author} {\bibinfo {author} {\bibfnamefont {S.}~\bibnamefont
  {Li}}, \bibinfo {author} {\bibfnamefont {Q.}~\bibnamefont {Zhang}}, \bibinfo
  {author} {\bibfnamefont {P.}~\bibnamefont {Ghaemi}}, \ and\ \bibinfo {author}
  {\bibfnamefont {M.~P.}\ \bibnamefont {Sarachik}},\ }\href {\doibase
  10.1103/PhysRevB.99.155302} {\bibfield  {journal} {\bibinfo  {journal} {Phys.
  Rev. B}\ }\textbf {\bibinfo {volume} {99}},\ \bibinfo {pages} {155302}
  (\bibinfo {year} {2019})}\BibitemShut {NoStop}%
\bibitem [{\citenamefont {Okamoto}\ \emph {et~al.}(1999)\citenamefont
  {Okamoto}, \citenamefont {Hosoya}, \citenamefont {Kawaji},\ and\
  \citenamefont {Yagi}}]{RN122}%
  \BibitemOpen
  \bibfield  {author} {\bibinfo {author} {\bibfnamefont {T.}~\bibnamefont
  {Okamoto}}, \bibinfo {author} {\bibfnamefont {K.}~\bibnamefont {Hosoya}},
  \bibinfo {author} {\bibfnamefont {S.}~\bibnamefont {Kawaji}}, \ and\ \bibinfo
  {author} {\bibfnamefont {A.}~\bibnamefont {Yagi}},\ }\href {\doibase
  10.1103/PhysRevLett.82.3875} {\bibfield  {journal} {\bibinfo  {journal}
  {Phys. Rev. Lett.}\ }\textbf {\bibinfo {volume} {82}},\ \bibinfo {pages}
  {3875} (\bibinfo {year} {1999})}\BibitemShut {NoStop}%
\bibitem [{\citenamefont {Tsui}\ \emph {et~al.}(2005)\citenamefont {Tsui},
  \citenamefont {Vitkalov}, \citenamefont {Sarachik},\ and\ \citenamefont
  {Klapwijk}}]{RN4}%
  \BibitemOpen
  \bibfield  {author} {\bibinfo {author} {\bibfnamefont {Y.}~\bibnamefont
  {Tsui}}, \bibinfo {author} {\bibfnamefont {S.~A.}\ \bibnamefont {Vitkalov}},
  \bibinfo {author} {\bibfnamefont {M.~P.}\ \bibnamefont {Sarachik}}, \ and\
  \bibinfo {author} {\bibfnamefont {T.~M.}\ \bibnamefont {Klapwijk}},\ }\href
  {\doibase 10.1103/PhysRevB.71.113308} {\bibfield  {journal} {\bibinfo
  {journal} {Phys. Rev. B}\ }\textbf {\bibinfo {volume} {71}},\ \bibinfo
  {pages} {113308} (\bibinfo {year} {2005})}\BibitemShut {NoStop}%
\bibitem [{\citenamefont {Young}\ \emph {et~al.}(2012)\citenamefont {Young},
  \citenamefont {Dean}, \citenamefont {Wang}, \citenamefont {Ren},
  \citenamefont {Cadden-Zimansky}, \citenamefont {Watanabe}, \citenamefont
  {Taniguchi}, \citenamefont {Hone}, \citenamefont {Shepard},\ and\
  \citenamefont {Kim}}]{NaturePhysics.8.7}%
  \BibitemOpen
  \bibfield  {author} {\bibinfo {author} {\bibfnamefont {A.~F.}\ \bibnamefont
  {Young}}, \bibinfo {author} {\bibfnamefont {C.~R.}\ \bibnamefont {Dean}},
  \bibinfo {author} {\bibfnamefont {L.}~\bibnamefont {Wang}}, \bibinfo {author}
  {\bibfnamefont {H.}~\bibnamefont {Ren}}, \bibinfo {author} {\bibfnamefont
  {P.}~\bibnamefont {Cadden-Zimansky}}, \bibinfo {author} {\bibfnamefont
  {K.}~\bibnamefont {Watanabe}}, \bibinfo {author} {\bibfnamefont
  {T.}~\bibnamefont {Taniguchi}}, \bibinfo {author} {\bibfnamefont
  {J.}~\bibnamefont {Hone}}, \bibinfo {author} {\bibfnamefont {K.~L.}\
  \bibnamefont {Shepard}}, \ and\ \bibinfo {author} {\bibfnamefont
  {P.}~\bibnamefont {Kim}},\ }\href {\doibase 10.1038/nphys2307} {\bibfield
  {journal} {\bibinfo  {journal} {Nature Physics}\ }\textbf {\bibinfo {volume}
  {8}},\ \bibinfo {pages} {550} (\bibinfo {year} {2012})}\BibitemShut {NoStop}%
\bibitem [{\citenamefont {Zhao}\ \emph {et~al.}(2012)\citenamefont {Zhao},
  \citenamefont {Cadden-Zimansky}, \citenamefont {Ghahari},\ and\ \citenamefont
  {Kim}}]{RN17}%
  \BibitemOpen
  \bibfield  {author} {\bibinfo {author} {\bibfnamefont {Y.}~\bibnamefont
  {Zhao}}, \bibinfo {author} {\bibfnamefont {P.}~\bibnamefont
  {Cadden-Zimansky}}, \bibinfo {author} {\bibfnamefont {F.}~\bibnamefont
  {Ghahari}}, \ and\ \bibinfo {author} {\bibfnamefont {P.}~\bibnamefont
  {Kim}},\ }\href {\doibase 10.1103/PhysRevLett.108.106804} {\bibfield
  {journal} {\bibinfo  {journal} {Phys. Rev. Lett.}\ }\textbf {\bibinfo
  {volume} {108}},\ \bibinfo {pages} {106804} (\bibinfo {year}
  {2012})}\BibitemShut {NoStop}%
\bibitem [{\citenamefont {Zeng}\ \emph {et~al.}(2019)\citenamefont {Zeng},
  \citenamefont {Li}, \citenamefont {Dietrich}, \citenamefont {Ghosh},
  \citenamefont {Watanabe}, \citenamefont {Taniguchi}, \citenamefont {Hone},\
  and\ \citenamefont {Dean}}]{RN18}%
  \BibitemOpen
  \bibfield  {author} {\bibinfo {author} {\bibfnamefont {Y.}~\bibnamefont
  {Zeng}}, \bibinfo {author} {\bibfnamefont {J.~I.~A.}\ \bibnamefont {Li}},
  \bibinfo {author} {\bibfnamefont {S.~A.}\ \bibnamefont {Dietrich}}, \bibinfo
  {author} {\bibfnamefont {O.~M.}\ \bibnamefont {Ghosh}}, \bibinfo {author}
  {\bibfnamefont {K.}~\bibnamefont {Watanabe}}, \bibinfo {author}
  {\bibfnamefont {T.}~\bibnamefont {Taniguchi}}, \bibinfo {author}
  {\bibfnamefont {J.}~\bibnamefont {Hone}}, \ and\ \bibinfo {author}
  {\bibfnamefont {C.~R.}\ \bibnamefont {Dean}},\ }\href {\doibase
  10.1103/PhysRevLett.122.137701} {\bibfield  {journal} {\bibinfo  {journal}
  {Phys. Rev. Lett.}\ }\textbf {\bibinfo {volume} {122}},\ \bibinfo {pages}
  {137701} (\bibinfo {year} {2019})}\BibitemShut {NoStop}%
\bibitem [{\citenamefont {Denteneer}\ and\ \citenamefont
  {Scalettar}(2003)}]{RN1}%
  \BibitemOpen
  \bibfield  {author} {\bibinfo {author} {\bibfnamefont {P.~J.~H.}\
  \bibnamefont {Denteneer}}\ and\ \bibinfo {author} {\bibfnamefont {R.~T.}\
  \bibnamefont {Scalettar}},\ }\href {\doibase 10.1103/PhysRevLett.90.246401}
  {\bibfield  {journal} {\bibinfo  {journal} {Phys. Rev. Lett.}\ }\textbf
  {\bibinfo {volume} {90}},\ \bibinfo {pages} {246401} (\bibinfo {year}
  {2003})}\BibitemShut {NoStop}%
\bibitem [{\citenamefont {Trivedi}\ \emph {et~al.}(2005)\citenamefont
  {Trivedi}, \citenamefont {Denteneer}, \citenamefont {Heidarian},\ and\
  \citenamefont {Scalettar}}]{RN3}%
  \BibitemOpen
  \bibfield  {author} {\bibinfo {author} {\bibfnamefont {N.}~\bibnamefont
  {Trivedi}}, \bibinfo {author} {\bibfnamefont {P.~J.~H.}\ \bibnamefont
  {Denteneer}}, \bibinfo {author} {\bibfnamefont {D.}~\bibnamefont
  {Heidarian}}, \ and\ \bibinfo {author} {\bibfnamefont {R.~T.}\ \bibnamefont
  {Scalettar}},\ }\href {\doibase 10.1007/BF02704167} {\bibfield  {journal}
  {\bibinfo  {journal} {Pramana}\ }\textbf {\bibinfo {volume} {64}},\ \bibinfo
  {pages} {1051} (\bibinfo {year} {2005})}\BibitemShut {NoStop}%
\bibitem [{\citenamefont {Hasan}\ and\ \citenamefont {Kane}(2010)}]{RN116}%
  \BibitemOpen
  \bibfield  {author} {\bibinfo {author} {\bibfnamefont {M.~Z.}\ \bibnamefont
  {Hasan}}\ and\ \bibinfo {author} {\bibfnamefont {C.~L.}\ \bibnamefont
  {Kane}},\ }\href {\doibase 10.1103/RevModPhys.82.3045} {\bibfield  {journal}
  {\bibinfo  {journal} {Rev. Mod. Phys.}\ }\textbf {\bibinfo {volume} {82}},\
  \bibinfo {pages} {3045} (\bibinfo {year} {2010})}\BibitemShut {NoStop}%
\bibitem [{\citenamefont {Zyuzin}\ and\ \citenamefont {Burkov}(2012)}]{RN120}%
  \BibitemOpen
  \bibfield  {author} {\bibinfo {author} {\bibfnamefont {A.~A.}\ \bibnamefont
  {Zyuzin}}\ and\ \bibinfo {author} {\bibfnamefont {A.~A.}\ \bibnamefont
  {Burkov}},\ }\href {\doibase 10.1103/PhysRevB.86.115133} {\bibfield
  {journal} {\bibinfo  {journal} {Phys. Rev. B}\ }\textbf {\bibinfo {volume}
  {86}},\ \bibinfo {pages} {115133} (\bibinfo {year} {2012})}\BibitemShut
  {NoStop}%
\bibitem [{\citenamefont {Geim}\ and\ \citenamefont {Novoselov}(2007)}]{RN108}%
  \BibitemOpen
  \bibfield  {author} {\bibinfo {author} {\bibfnamefont {A.~K.}\ \bibnamefont
  {Geim}}\ and\ \bibinfo {author} {\bibfnamefont {K.~S.}\ \bibnamefont
  {Novoselov}},\ }\href {\doibase 10.1038/nmat1849} {\bibfield  {journal}
  {\bibinfo  {journal} {Nat. Mater.}\ }\textbf {\bibinfo {volume} {6}},\
  \bibinfo {pages} {183} (\bibinfo {year} {2007})}\BibitemShut {NoStop}%
\bibitem [{\citenamefont {Gupta}\ \emph {et~al.}(2015)\citenamefont {Gupta},
  \citenamefont {Sakthivel},\ and\ \citenamefont {Seal}}]{RN109}%
  \BibitemOpen
  \bibfield  {author} {\bibinfo {author} {\bibfnamefont {A.}~\bibnamefont
  {Gupta}}, \bibinfo {author} {\bibfnamefont {T.}~\bibnamefont {Sakthivel}}, \
  and\ \bibinfo {author} {\bibfnamefont {S.}~\bibnamefont {Seal}},\ }\href
  {\doibase https://doi.org/10.1016/j.pmatsci.2015.02.002} {\bibfield
  {journal} {\bibinfo  {journal} {Prog. Mater. Sci.}\ }\textbf {\bibinfo
  {volume} {73}},\ \bibinfo {pages} {44 } (\bibinfo {year} {2015})}\BibitemShut
  {NoStop}%
\bibitem [{\citenamefont {Li}\ \emph {et~al.}(2018)\citenamefont {Li},
  \citenamefont {Zhang}, \citenamefont {li}, \citenamefont {Wang},\ and\
  \citenamefont {Qin}}]{LI2018845}%
  \BibitemOpen
  \bibfield  {author} {\bibinfo {author} {\bibfnamefont {Z.}~\bibnamefont
  {Li}}, \bibinfo {author} {\bibfnamefont {W.}~\bibnamefont {Zhang}}, \bibinfo
  {author} {\bibfnamefont {Y.}~\bibnamefont {li}}, \bibinfo {author}
  {\bibfnamefont {H.}~\bibnamefont {Wang}}, \ and\ \bibinfo {author}
  {\bibfnamefont {Z.}~\bibnamefont {Qin}},\ }\href {\doibase
  https://doi.org/10.1016/j.cej.2017.10.023} {\bibfield  {journal} {\bibinfo
  {journal} {Chemical Engineering Journal}\ }\textbf {\bibinfo {volume}
  {334}},\ \bibinfo {pages} {845 } (\bibinfo {year} {2018})}\BibitemShut
  {NoStop}%
\bibitem [{\citenamefont {Chen}\ \emph {et~al.}(2008)\citenamefont {Chen},
  \citenamefont {Müller}, \citenamefont {Gilmore}, \citenamefont {Wallace},\
  and\ \citenamefont {Li}}]{RN113}%
  \BibitemOpen
  \bibfield  {author} {\bibinfo {author} {\bibfnamefont {H.}~\bibnamefont
  {Chen}}, \bibinfo {author} {\bibfnamefont {M.~B.}\ \bibnamefont {Müller}},
  \bibinfo {author} {\bibfnamefont {K.~J.}\ \bibnamefont {Gilmore}}, \bibinfo
  {author} {\bibfnamefont {G.~G.}\ \bibnamefont {Wallace}}, \ and\ \bibinfo
  {author} {\bibfnamefont {D.}~\bibnamefont {Li}},\ }\href {\doibase
  10.1002/adma.200800757} {\bibfield  {journal} {\bibinfo  {journal} {Advanced
  Materials}\ }\textbf {\bibinfo {volume} {20}},\ \bibinfo {pages} {3557}
  (\bibinfo {year} {2008})}\BibitemShut {NoStop}%
\bibitem [{\citenamefont {Horng}\ \emph {et~al.}(2011)\citenamefont {Horng},
  \citenamefont {Chen}, \citenamefont {Geng}, \citenamefont {Girit},
  \citenamefont {Zhang}, \citenamefont {Hao}, \citenamefont {Bechtel},
  \citenamefont {Martin}, \citenamefont {Zettl}, \citenamefont {Crommie},
  \citenamefont {Shen},\ and\ \citenamefont {Wang}}]{RN114}%
  \BibitemOpen
  \bibfield  {author} {\bibinfo {author} {\bibfnamefont {J.}~\bibnamefont
  {Horng}}, \bibinfo {author} {\bibfnamefont {C.-F.}\ \bibnamefont {Chen}},
  \bibinfo {author} {\bibfnamefont {B.}~\bibnamefont {Geng}}, \bibinfo {author}
  {\bibfnamefont {C.}~\bibnamefont {Girit}}, \bibinfo {author} {\bibfnamefont
  {Y.}~\bibnamefont {Zhang}}, \bibinfo {author} {\bibfnamefont
  {Z.}~\bibnamefont {Hao}}, \bibinfo {author} {\bibfnamefont {H.~A.}\
  \bibnamefont {Bechtel}}, \bibinfo {author} {\bibfnamefont {M.}~\bibnamefont
  {Martin}}, \bibinfo {author} {\bibfnamefont {A.}~\bibnamefont {Zettl}},
  \bibinfo {author} {\bibfnamefont {M.~F.}\ \bibnamefont {Crommie}}, \bibinfo
  {author} {\bibfnamefont {Y.~R.}\ \bibnamefont {Shen}}, \ and\ \bibinfo
  {author} {\bibfnamefont {F.}~\bibnamefont {Wang}},\ }\href {\doibase
  10.1103/PhysRevB.83.165113} {\bibfield  {journal} {\bibinfo  {journal} {Phys.
  Rev. B}\ }\textbf {\bibinfo {volume} {83}},\ \bibinfo {pages} {165113}
  (\bibinfo {year} {2011})}\BibitemShut {NoStop}%
\bibitem [{\citenamefont {Caldas}\ and\ \citenamefont {Ramos}(2009)}]{RN6}%
  \BibitemOpen
  \bibfield  {author} {\bibinfo {author} {\bibfnamefont {H.}~\bibnamefont
  {Caldas}}\ and\ \bibinfo {author} {\bibfnamefont {R.~O.}\ \bibnamefont
  {Ramos}},\ }\href {\doibase 10.1103/PhysRevB.80.115428} {\bibfield  {journal}
  {\bibinfo  {journal} {Phys. Rev. B}\ }\textbf {\bibinfo {volume} {80}},\
  \bibinfo {pages} {115428} (\bibinfo {year} {2009})}\BibitemShut {NoStop}%
\bibitem [{\citenamefont {Hwang}\ and\ \citenamefont {Das~Sarma}(2009)}]{RN2}%
  \BibitemOpen
  \bibfield  {author} {\bibinfo {author} {\bibfnamefont {E.~H.}\ \bibnamefont
  {Hwang}}\ and\ \bibinfo {author} {\bibfnamefont {S.}~\bibnamefont
  {Das~Sarma}},\ }\href {\doibase 10.1103/PhysRevB.80.075417} {\bibfield
  {journal} {\bibinfo  {journal} {Phys. Rev. B}\ }\textbf {\bibinfo {volume}
  {80}},\ \bibinfo {pages} {075417} (\bibinfo {year} {2009})}\BibitemShut
  {NoStop}%
\bibitem [{\citenamefont {Wehling}\ \emph {et~al.}(2011)\citenamefont
  {Wehling}, \citenamefont {\ifmmode \mbox{\c{S}}\else \c{S}\fi{}a\ifmmode
  \mbox{\c{s}}\else \c{s}\fi{}\ifmmode \imath \else \i
  \fi{}o\ifmmode~\breve{g}\else \u{g}\fi{}lu}, \citenamefont {Friedrich},
  \citenamefont {Lichtenstein}, \citenamefont {Katsnelson},\ and\ \citenamefont
  {Bl\"ugel}}]{PhysRevLett.106.236805}%
  \BibitemOpen
  \bibfield  {author} {\bibinfo {author} {\bibfnamefont {T.~O.}\ \bibnamefont
  {Wehling}}, \bibinfo {author} {\bibfnamefont {E.}~\bibnamefont {\ifmmode
  \mbox{\c{S}}\else \c{S}\fi{}a\ifmmode \mbox{\c{s}}\else \c{s}\fi{}\ifmmode
  \imath \else \i \fi{}o\ifmmode~\breve{g}\else \u{g}\fi{}lu}}, \bibinfo
  {author} {\bibfnamefont {C.}~\bibnamefont {Friedrich}}, \bibinfo {author}
  {\bibfnamefont {A.~I.}\ \bibnamefont {Lichtenstein}}, \bibinfo {author}
  {\bibfnamefont {M.~I.}\ \bibnamefont {Katsnelson}}, \ and\ \bibinfo {author}
  {\bibfnamefont {S.}~\bibnamefont {Bl\"ugel}},\ }\href {\doibase
  10.1103/PhysRevLett.106.236805} {\bibfield  {journal} {\bibinfo  {journal}
  {Phys. Rev. Lett.}\ }\textbf {\bibinfo {volume} {106}},\ \bibinfo {pages}
  {236805} (\bibinfo {year} {2011})}\BibitemShut {NoStop}%
\bibitem [{\citenamefont {Tang}\ \emph {et~al.}(2018)\citenamefont {Tang},
  \citenamefont {Leaw}, \citenamefont {Rodrigues}, \citenamefont {Herbut},
  \citenamefont {Sengupta}, \citenamefont {Assaad},\ and\ \citenamefont
  {Adam}}]{Tang2018}%
  \BibitemOpen
  \bibfield  {author} {\bibinfo {author} {\bibfnamefont {H.-K.}\ \bibnamefont
  {Tang}}, \bibinfo {author} {\bibfnamefont {J.~N.}\ \bibnamefont {Leaw}},
  \bibinfo {author} {\bibfnamefont {J.~N.~B.}\ \bibnamefont {Rodrigues}},
  \bibinfo {author} {\bibfnamefont {I.~F.}\ \bibnamefont {Herbut}}, \bibinfo
  {author} {\bibfnamefont {P.}~\bibnamefont {Sengupta}}, \bibinfo {author}
  {\bibfnamefont {F.~F.}\ \bibnamefont {Assaad}}, \ and\ \bibinfo {author}
  {\bibfnamefont {S.}~\bibnamefont {Adam}},\ }\href {\doibase
  10.1126/science.aao2934} {\bibfield  {journal} {\bibinfo  {journal}
  {Science}\ }\textbf {\bibinfo {volume} {361}},\ \bibinfo {pages} {570}
  (\bibinfo {year} {2018})}\BibitemShut {NoStop}%
\bibitem [{\citenamefont {Hesselmann}\ \emph {et~al.}(2019)\citenamefont
  {Hesselmann}, \citenamefont {Lang}, \citenamefont {Schuler}, \citenamefont
  {Wessel},\ and\ \citenamefont {L{\"a}uchli}}]{Hesselmanneaav6869}%
  \BibitemOpen
  \bibfield  {author} {\bibinfo {author} {\bibfnamefont {S.}~\bibnamefont
  {Hesselmann}}, \bibinfo {author} {\bibfnamefont {T.~C.}\ \bibnamefont
  {Lang}}, \bibinfo {author} {\bibfnamefont {M.}~\bibnamefont {Schuler}},
  \bibinfo {author} {\bibfnamefont {S.}~\bibnamefont {Wessel}}, \ and\ \bibinfo
  {author} {\bibfnamefont {A.~M.}\ \bibnamefont {L{\"a}uchli}},\ }\href
  {\doibase 10.1126/science.aav6869} {\bibfield  {journal} {\bibinfo  {journal}
  {Science}\ }\textbf {\bibinfo {volume} {366}},\ \bibinfo {pages} {eaav6869}
  (\bibinfo {year} {2019})}\BibitemShut {NoStop}%
\bibitem [{\citenamefont {Pereira}\ \emph {et~al.}(2008)\citenamefont
  {Pereira}, \citenamefont {Lopes~dos Santos},\ and\ \citenamefont
  {Castro~Neto}}]{PhysRevB.77.115109}%
  \BibitemOpen
  \bibfield  {author} {\bibinfo {author} {\bibfnamefont {V.~M.}\ \bibnamefont
  {Pereira}}, \bibinfo {author} {\bibfnamefont {J.~M.~B.}\ \bibnamefont
  {Lopes~dos Santos}}, \ and\ \bibinfo {author} {\bibfnamefont {A.~H.}\
  \bibnamefont {Castro~Neto}},\ }\href {\doibase 10.1103/PhysRevB.77.115109}
  {\bibfield  {journal} {\bibinfo  {journal} {Phys. Rev. B}\ }\textbf {\bibinfo
  {volume} {77}},\ \bibinfo {pages} {115109} (\bibinfo {year}
  {2008})}\BibitemShut {NoStop}%
\bibitem [{\citenamefont {Ma}\ \emph {et~al.}(2018)\citenamefont {Ma},
  \citenamefont {Zhang}, \citenamefont {Chang}, \citenamefont {Hung},\ and\
  \citenamefont {Scalettar}}]{RN5}%
  \BibitemOpen
  \bibfield  {author} {\bibinfo {author} {\bibfnamefont {T.}~\bibnamefont
  {Ma}}, \bibinfo {author} {\bibfnamefont {L.}~\bibnamefont {Zhang}}, \bibinfo
  {author} {\bibfnamefont {C.-C.}\ \bibnamefont {Chang}}, \bibinfo {author}
  {\bibfnamefont {H.-H.}\ \bibnamefont {Hung}}, \ and\ \bibinfo {author}
  {\bibfnamefont {R.~T.}\ \bibnamefont {Scalettar}},\ }\href {\doibase
  10.1103/PhysRevLett.120.116601} {\bibfield  {journal} {\bibinfo  {journal}
  {Phys. Rev. Lett.}\ }\textbf {\bibinfo {volume} {120}},\ \bibinfo {pages}
  {116601} (\bibinfo {year} {2018})}\BibitemShut {NoStop}%
\bibitem [{\citenamefont {Goerbig}(2011)}]{RN124}%
  \BibitemOpen
  \bibfield  {author} {\bibinfo {author} {\bibfnamefont {M.~O.}\ \bibnamefont
  {Goerbig}},\ }\href {\doibase 10.1103/RevModPhys.83.1193} {\bibfield
  {journal} {\bibinfo  {journal} {Rev. Mod. Phys.}\ }\textbf {\bibinfo {volume}
  {83}},\ \bibinfo {pages} {1193} (\bibinfo {year} {2011})}\BibitemShut
  {NoStop}%
\bibitem [{\citenamefont {Barlas}\ \emph {et~al.}(2012)\citenamefont {Barlas},
  \citenamefont {Yang},\ and\ \citenamefont {MacDonald}}]{RN125}%
  \BibitemOpen
  \bibfield  {author} {\bibinfo {author} {\bibfnamefont {Y.}~\bibnamefont
  {Barlas}}, \bibinfo {author} {\bibfnamefont {K.}~\bibnamefont {Yang}}, \ and\
  \bibinfo {author} {\bibfnamefont {A.~H.}\ \bibnamefont {MacDonald}},\ }\href
  {\doibase 10.1088/0957-4484/23/5/052001} {\bibfield  {journal} {\bibinfo
  {journal} {Nanotechnology}\ }\textbf {\bibinfo {volume} {23}},\ \bibinfo
  {pages} {052001} (\bibinfo {year} {2012})}\BibitemShut {NoStop}%
\bibitem [{\citenamefont {Papadakis}\ \emph {et~al.}(2000)\citenamefont
  {Papadakis}, \citenamefont {De~Poortere}, \citenamefont {Shayegan},\ and\
  \citenamefont {Winkler}}]{PhysRevLett.84.5592}%
  \BibitemOpen
  \bibfield  {author} {\bibinfo {author} {\bibfnamefont {S.~J.}\ \bibnamefont
  {Papadakis}}, \bibinfo {author} {\bibfnamefont {E.~P.}\ \bibnamefont
  {De~Poortere}}, \bibinfo {author} {\bibfnamefont {M.}~\bibnamefont
  {Shayegan}}, \ and\ \bibinfo {author} {\bibfnamefont {R.}~\bibnamefont
  {Winkler}},\ }\href {\doibase 10.1103/PhysRevLett.84.5592} {\bibfield
  {journal} {\bibinfo  {journal} {Phys. Rev. Lett.}\ }\textbf {\bibinfo
  {volume} {84}},\ \bibinfo {pages} {5592} (\bibinfo {year}
  {2000})}\BibitemShut {NoStop}%
\bibitem [{\citenamefont {Antipov}\ \emph {et~al.}(2016)\citenamefont
  {Antipov}, \citenamefont {Javanmard}, \citenamefont {Ribeiro},\ and\
  \citenamefont {Kirchner}}]{PhysRevLett.117.146601}%
  \BibitemOpen
  \bibfield  {author} {\bibinfo {author} {\bibfnamefont {A.~E.}\ \bibnamefont
  {Antipov}}, \bibinfo {author} {\bibfnamefont {Y.}~\bibnamefont {Javanmard}},
  \bibinfo {author} {\bibfnamefont {P.}~\bibnamefont {Ribeiro}}, \ and\
  \bibinfo {author} {\bibfnamefont {S.}~\bibnamefont {Kirchner}},\ }\href
  {\doibase 10.1103/PhysRevLett.117.146601} {\bibfield  {journal} {\bibinfo
  {journal} {Phys. Rev. Lett.}\ }\textbf {\bibinfo {volume} {117}},\ \bibinfo
  {pages} {146601} (\bibinfo {year} {2016})}\BibitemShut {NoStop}%
\bibitem [{\citenamefont {Denteneer}\ \emph {et~al.}(1999)\citenamefont
  {Denteneer}, \citenamefont {Scalettar},\ and\ \citenamefont
  {Trivedi}}]{RN101}%
  \BibitemOpen
  \bibfield  {author} {\bibinfo {author} {\bibfnamefont {P.~J.~H.}\
  \bibnamefont {Denteneer}}, \bibinfo {author} {\bibfnamefont {R.~T.}\
  \bibnamefont {Scalettar}}, \ and\ \bibinfo {author} {\bibfnamefont
  {N.}~\bibnamefont {Trivedi}},\ }\href {\doibase 10.1103/PhysRevLett.83.4610}
  {\bibfield  {journal} {\bibinfo  {journal} {Phys. Rev. Lett.}\ }\textbf
  {\bibinfo {volume} {83}},\ \bibinfo {pages} {4610} (\bibinfo {year}
  {1999})}\BibitemShut {NoStop}%
\bibitem [{\citenamefont {White}\ \emph {et~al.}(1989)\citenamefont {White},
  \citenamefont {Scalapino}, \citenamefont {Sugar}, \citenamefont {Loh},
  \citenamefont {Gubernatis},\ and\ \citenamefont {Scalettar}}]{RN102}%
  \BibitemOpen
  \bibfield  {author} {\bibinfo {author} {\bibfnamefont {S.~R.}\ \bibnamefont
  {White}}, \bibinfo {author} {\bibfnamefont {D.~J.}\ \bibnamefont
  {Scalapino}}, \bibinfo {author} {\bibfnamefont {R.~L.}\ \bibnamefont
  {Sugar}}, \bibinfo {author} {\bibfnamefont {E.~Y.}\ \bibnamefont {Loh}},
  \bibinfo {author} {\bibfnamefont {J.~E.}\ \bibnamefont {Gubernatis}}, \ and\
  \bibinfo {author} {\bibfnamefont {R.~T.}\ \bibnamefont {Scalettar}},\ }\href
  {\doibase 10.1103/PhysRevB.40.506} {\bibfield  {journal} {\bibinfo  {journal}
  {Phys. Rev. B}\ }\textbf {\bibinfo {volume} {40}},\ \bibinfo {pages} {506}
  (\bibinfo {year} {1989})}\BibitemShut {NoStop}%
\bibitem [{\citenamefont {Paiva}\ \emph {et~al.}(2015)\citenamefont {Paiva},
  \citenamefont {Khatami}, \citenamefont {Yang}, \citenamefont {Rousseau},
  \citenamefont {Jarrell}, \citenamefont {Moreno}, \citenamefont {Hulet},\ and\
  \citenamefont {Scalettar}}]{RN115}%
  \BibitemOpen
  \bibfield  {author} {\bibinfo {author} {\bibfnamefont {T.}~\bibnamefont
  {Paiva}}, \bibinfo {author} {\bibfnamefont {E.}~\bibnamefont {Khatami}},
  \bibinfo {author} {\bibfnamefont {S.}~\bibnamefont {Yang}}, \bibinfo {author}
  {\bibfnamefont {V.}~\bibnamefont {Rousseau}}, \bibinfo {author}
  {\bibfnamefont {M.}~\bibnamefont {Jarrell}}, \bibinfo {author} {\bibfnamefont
  {J.}~\bibnamefont {Moreno}}, \bibinfo {author} {\bibfnamefont {R.~G.}\
  \bibnamefont {Hulet}}, \ and\ \bibinfo {author} {\bibfnamefont {R.~T.}\
  \bibnamefont {Scalettar}},\ }\href {\doibase 10.1103/PhysRevLett.115.240402}
  {\bibfield  {journal} {\bibinfo  {journal} {Phys. Rev. Lett.}\ }\textbf
  {\bibinfo {volume} {115}},\ \bibinfo {pages} {240402} (\bibinfo {year}
  {2015})}\BibitemShut {NoStop}%
\bibitem [{\citenamefont {Trivedi}\ \emph
  {et~al.}(1996{\natexlab{a}})\citenamefont {Trivedi}, \citenamefont
  {Scalettar},\ and\ \citenamefont {Randeria}}]{RN103a}%
  \BibitemOpen
  \bibfield  {author} {\bibinfo {author} {\bibfnamefont {N.}~\bibnamefont
  {Trivedi}}, \bibinfo {author} {\bibfnamefont {R.~T.}\ \bibnamefont
  {Scalettar}}, \ and\ \bibinfo {author} {\bibfnamefont {M.}~\bibnamefont
  {Randeria}},\ }\href {\doibase 10.1103/PhysRevB.54.R3756} {\bibfield
  {journal} {\bibinfo  {journal} {Phys. Rev. B}\ }\textbf {\bibinfo {volume}
  {54}},\ \bibinfo {pages} {R3756} (\bibinfo {year}
  {1996}{\natexlab{a}})}\BibitemShut {NoStop}%
\bibitem [{\citenamefont {Lee}\ \emph {et~al.}(2007)\citenamefont {Lee},
  \citenamefont {Kune\ifmmode~\check{s}\else \v{s}\fi{}}, \citenamefont
  {Scalettar},\ and\ \citenamefont {Pickett}}]{RN103b}%
  \BibitemOpen
  \bibfield  {author} {\bibinfo {author} {\bibfnamefont {K.-W.}\ \bibnamefont
  {Lee}}, \bibinfo {author} {\bibfnamefont {J.}~\bibnamefont
  {Kune\ifmmode~\check{s}\else \v{s}\fi{}}}, \bibinfo {author} {\bibfnamefont
  {R.~T.}\ \bibnamefont {Scalettar}}, \ and\ \bibinfo {author} {\bibfnamefont
  {W.~E.}\ \bibnamefont {Pickett}},\ }\href {\doibase
  10.1103/PhysRevB.76.144513} {\bibfield  {journal} {\bibinfo  {journal} {Phys.
  Rev. B}\ }\textbf {\bibinfo {volume} {76}},\ \bibinfo {pages} {144513}
  (\bibinfo {year} {2007})}\BibitemShut {NoStop}%
\bibitem [{\citenamefont {Scalapino}\ \emph {et~al.}(1993)\citenamefont
  {Scalapino}, \citenamefont {White},\ and\ \citenamefont {Zhang}}]{RN103c}%
  \BibitemOpen
  \bibfield  {author} {\bibinfo {author} {\bibfnamefont {D.~J.}\ \bibnamefont
  {Scalapino}}, \bibinfo {author} {\bibfnamefont {S.~R.}\ \bibnamefont
  {White}}, \ and\ \bibinfo {author} {\bibfnamefont {S.}~\bibnamefont
  {Zhang}},\ }\href {\doibase 10.1103/PhysRevB.47.7995} {\bibfield  {journal}
  {\bibinfo  {journal} {Phys. Rev. B}\ }\textbf {\bibinfo {volume} {47}},\
  \bibinfo {pages} {7995} (\bibinfo {year} {1993})}\BibitemShut {NoStop}%
\bibitem [{\citenamefont {Pathria}\ and\ \citenamefont {Beale}(2011)}]{RN103d}%
  \BibitemOpen
  \bibfield  {author} {\bibinfo {author} {\bibfnamefont {R.}~\bibnamefont
  {Pathria}}\ and\ \bibinfo {author} {\bibfnamefont {P.~D.}\ \bibnamefont
  {Beale}},\ }in\ \href {\doibase
  https://doi.org/10.1016/B978-0-12-382188-1.00008-6} {\emph {\bibinfo
  {booktitle} {Statistical Mechanics (Third Edition)}}},\ \bibinfo {editor}
  {edited by\ \bibinfo {editor} {\bibfnamefont {R.}~\bibnamefont {Pathria}}\
  and\ \bibinfo {editor} {\bibfnamefont {P.~D.}\ \bibnamefont {Beale}}}\
  (\bibinfo  {publisher} {Academic Press},\ \bibinfo {address} {Boston},\
  \bibinfo {year} {2011})\ \bibinfo {edition} {third edition}\ ed.,\ pp.\
  \bibinfo {pages} {231 -- 273}\BibitemShut {NoStop}%
\bibitem [{\citenamefont {Trivedi}\ \emph
  {et~al.}(1996{\natexlab{b}})\citenamefont {Trivedi}, \citenamefont
  {Scalettar},\ and\ \citenamefont {Randeria}}]{Trivedi1996}%
  \BibitemOpen
  \bibfield  {author} {\bibinfo {author} {\bibfnamefont {N.}~\bibnamefont
  {Trivedi}}, \bibinfo {author} {\bibfnamefont {R.~T.}\ \bibnamefont
  {Scalettar}}, \ and\ \bibinfo {author} {\bibfnamefont {M.}~\bibnamefont
  {Randeria}},\ }\href {\doibase 10.1103/PhysRevB.54.R3756} {\bibfield
  {journal} {\bibinfo  {journal} {Phys. Rev. B}\ }\textbf {\bibinfo {volume}
  {54}},\ \bibinfo {pages} {R3756} (\bibinfo {year}
  {1996}{\natexlab{b}})}\BibitemShut {NoStop}%
\bibitem [{\citenamefont {Scalettar}\ \emph {et~al.}(1999)\citenamefont
  {Scalettar}, \citenamefont {Trivedi},\ and\ \citenamefont
  {Huscroft}}]{Scalettar1999}%
  \BibitemOpen
  \bibfield  {author} {\bibinfo {author} {\bibfnamefont {R.~T.}\ \bibnamefont
  {Scalettar}}, \bibinfo {author} {\bibfnamefont {N.}~\bibnamefont {Trivedi}},
  \ and\ \bibinfo {author} {\bibfnamefont {C.}~\bibnamefont {Huscroft}},\
  }\href {\doibase 10.1103/PhysRevB.59.4364} {\bibfield  {journal} {\bibinfo
  {journal} {Phys. Rev. B}\ }\textbf {\bibinfo {volume} {59}},\ \bibinfo
  {pages} {4364} (\bibinfo {year} {1999})}\BibitemShut {NoStop}%
\bibitem [{\citenamefont {Mondaini}\ \emph {et~al.}(2012)\citenamefont
  {Mondaini}, \citenamefont {Bouadim}, \citenamefont {Paiva},\ and\
  \citenamefont {dos Santos}}]{Mondaini2012}%
  \BibitemOpen
  \bibfield  {author} {\bibinfo {author} {\bibfnamefont {R.}~\bibnamefont
  {Mondaini}}, \bibinfo {author} {\bibfnamefont {K.}~\bibnamefont {Bouadim}},
  \bibinfo {author} {\bibfnamefont {T.}~\bibnamefont {Paiva}}, \ and\ \bibinfo
  {author} {\bibfnamefont {R.~R.}\ \bibnamefont {dos Santos}},\ }\href
  {\doibase 10.1103/PhysRevB.85.125127} {\bibfield  {journal} {\bibinfo
  {journal} {Phys. Rev. B}\ }\textbf {\bibinfo {volume} {85}},\ \bibinfo
  {pages} {125127} (\bibinfo {year} {2012})}\BibitemShut {NoStop}%
\bibitem [{\citenamefont {Lederer}\ \emph {et~al.}(2017)\citenamefont
  {Lederer}, \citenamefont {Schattner}, \citenamefont {Berg},\ and\
  \citenamefont {Kivelson}}]{Lederer2017}%
  \BibitemOpen
  \bibfield  {author} {\bibinfo {author} {\bibfnamefont {S.}~\bibnamefont
  {Lederer}}, \bibinfo {author} {\bibfnamefont {Y.}~\bibnamefont {Schattner}},
  \bibinfo {author} {\bibfnamefont {E.}~\bibnamefont {Berg}}, \ and\ \bibinfo
  {author} {\bibfnamefont {S.~A.}\ \bibnamefont {Kivelson}},\ }\href {\doibase
  10.1073/pnas.1620651114} {\bibfield  {journal} {\bibinfo  {journal} {Proc
  Natl Acad Sci}\ }\textbf {\bibinfo {volume} {114}},\ \bibinfo {pages} {4905}
  (\bibinfo {year} {2017})}\BibitemShut {NoStop}%
\bibitem [{\citenamefont {Huang}\ \emph {et~al.}(2019)\citenamefont {Huang},
  \citenamefont {Sheppard}, \citenamefont {Moritz},\ and\ \citenamefont
  {Devereaux}}]{Huang2019}%
  \BibitemOpen
  \bibfield  {author} {\bibinfo {author} {\bibfnamefont {E.~W.}\ \bibnamefont
  {Huang}}, \bibinfo {author} {\bibfnamefont {R.}~\bibnamefont {Sheppard}},
  \bibinfo {author} {\bibfnamefont {B.}~\bibnamefont {Moritz}}, \ and\ \bibinfo
  {author} {\bibfnamefont {T.~P.}\ \bibnamefont {Devereaux}},\ }\href {\doibase
  10.1126/science.aau7063} {\bibfield  {journal} {\bibinfo  {journal}
  {Science}\ }\textbf {\bibinfo {volume} {366}},\ \bibinfo {pages} {987}
  (\bibinfo {year} {2019})}\BibitemShut {NoStop}%
\bibitem [{\citenamefont {Trivedi}\ and\ \citenamefont
  {Randeria}(1995)}]{RN104}%
  \BibitemOpen
  \bibfield  {author} {\bibinfo {author} {\bibfnamefont {N.}~\bibnamefont
  {Trivedi}}\ and\ \bibinfo {author} {\bibfnamefont {M.}~\bibnamefont
  {Randeria}},\ }\href {\doibase 10.1103/PhysRevLett.75.312} {\bibfield
  {journal} {\bibinfo  {journal} {Phys. Rev. Lett.}\ }\textbf {\bibinfo
  {volume} {75}},\ \bibinfo {pages} {312} (\bibinfo {year} {1995})}\BibitemShut
  {NoStop}%
\bibitem [{\citenamefont {Guillemette}\ \emph {et~al.}(2018)\citenamefont
  {Guillemette}, \citenamefont {Hemsworth}, \citenamefont {Vlasov},
  \citenamefont {Kirman}, \citenamefont {Mahvash}, \citenamefont {L\'evesque},
  \citenamefont {Siaj}, \citenamefont {Martel}, \citenamefont {Gervais},
  \citenamefont {Studenikin}, \citenamefont {Sachrajda},\ and\ \citenamefont
  {Szkopek}}]{PhysRevB.97.161402}%
  \BibitemOpen
  \bibfield  {author} {\bibinfo {author} {\bibfnamefont {J.}~\bibnamefont
  {Guillemette}}, \bibinfo {author} {\bibfnamefont {N.}~\bibnamefont
  {Hemsworth}}, \bibinfo {author} {\bibfnamefont {A.}~\bibnamefont {Vlasov}},
  \bibinfo {author} {\bibfnamefont {J.}~\bibnamefont {Kirman}}, \bibinfo
  {author} {\bibfnamefont {F.}~\bibnamefont {Mahvash}}, \bibinfo {author}
  {\bibfnamefont {P.~L.}\ \bibnamefont {L\'evesque}}, \bibinfo {author}
  {\bibfnamefont {M.}~\bibnamefont {Siaj}}, \bibinfo {author} {\bibfnamefont
  {R.}~\bibnamefont {Martel}}, \bibinfo {author} {\bibfnamefont
  {G.}~\bibnamefont {Gervais}}, \bibinfo {author} {\bibfnamefont
  {S.}~\bibnamefont {Studenikin}}, \bibinfo {author} {\bibfnamefont
  {A.}~\bibnamefont {Sachrajda}}, \ and\ \bibinfo {author} {\bibfnamefont
  {T.}~\bibnamefont {Szkopek}},\ }\href {\doibase 10.1103/PhysRevB.97.161402}
  {\bibfield  {journal} {\bibinfo  {journal} {Phys. Rev. B}\ }\textbf {\bibinfo
  {volume} {97}},\ \bibinfo {pages} {161402} (\bibinfo {year}
  {2018})}\BibitemShut {NoStop}%
\bibitem [{\citenamefont {Sch\"uler}\ \emph {et~al.}(2013)\citenamefont
  {Sch\"uler}, \citenamefont {R\"osner}, \citenamefont {Wehling}, \citenamefont
  {Lichtenstein},\ and\ \citenamefont {Katsnelson}}]{PhysRevLett.111.036601}%
  \BibitemOpen
  \bibfield  {author} {\bibinfo {author} {\bibfnamefont {M.}~\bibnamefont
  {Sch\"uler}}, \bibinfo {author} {\bibfnamefont {M.}~\bibnamefont {R\"osner}},
  \bibinfo {author} {\bibfnamefont {T.~O.}\ \bibnamefont {Wehling}}, \bibinfo
  {author} {\bibfnamefont {A.~I.}\ \bibnamefont {Lichtenstein}}, \ and\
  \bibinfo {author} {\bibfnamefont {M.~I.}\ \bibnamefont {Katsnelson}},\ }\href
  {\doibase 10.1103/PhysRevLett.111.036601} {\bibfield  {journal} {\bibinfo
  {journal} {Phys. Rev. Lett.}\ }\textbf {\bibinfo {volume} {111}},\ \bibinfo
  {pages} {036601} (\bibinfo {year} {2013})}\BibitemShut {NoStop}%
\bibitem [{\citenamefont {Matveev}\ \emph {et~al.}(1995)\citenamefont
  {Matveev}, \citenamefont {Glazman}, \citenamefont {Clarke}, \citenamefont
  {Ephron},\ and\ \citenamefont {Beasley}}]{PhysRevB.52.5289}%
  \BibitemOpen
  \bibfield  {author} {\bibinfo {author} {\bibfnamefont {K.~A.}\ \bibnamefont
  {Matveev}}, \bibinfo {author} {\bibfnamefont {L.~I.}\ \bibnamefont
  {Glazman}}, \bibinfo {author} {\bibfnamefont {P.}~\bibnamefont {Clarke}},
  \bibinfo {author} {\bibfnamefont {D.}~\bibnamefont {Ephron}}, \ and\ \bibinfo
  {author} {\bibfnamefont {M.~R.}\ \bibnamefont {Beasley}},\ }\href {\doibase
  10.1103/PhysRevB.52.5289} {\bibfield  {journal} {\bibinfo  {journal} {Phys.
  Rev. B}\ }\textbf {\bibinfo {volume} {52}},\ \bibinfo {pages} {5289}
  (\bibinfo {year} {1995})}\BibitemShut {NoStop}%
\bibitem [{\citenamefont {{Van Cong}}(2004)}]{RN11}%
  \BibitemOpen
  \bibfield  {author} {\bibinfo {author} {\bibfnamefont {H.}~\bibnamefont {{Van
  Cong}}},\ }\href {\doibase https://doi.org/10.1016/j.physe.2003.12.125}
  {\bibfield  {journal} {\bibinfo  {journal} {Physica E}\ }\textbf {\bibinfo
  {volume} {22}},\ \bibinfo {pages} {924 } (\bibinfo {year}
  {2004})}\BibitemShut {NoStop}%
\bibitem [{Note1()}]{Note1}%
  \BibitemOpen
  \bibinfo {note} {Note that this value is fairly close to the ones obtained
  directly at the ground state for much larger lattices~\cite {Sorella2012},
  attesting the overall small finite size effects in the
  conductivity.}\BibitemShut {Stop}%
\bibitem [{\citenamefont {Rhodes}(2019)}]{NatureMaterials2019.18.6}%
  \BibitemOpen
  \bibfield  {author} {\bibinfo {author} {\bibfnamefont {D.}~\bibnamefont
  {Rhodes}},\ }\href {\doibase 10.1038/s41563-019-0366-8} {\bibfield  {journal}
  {\bibinfo  {journal} {Nat. Mater.}\ }\textbf {\bibinfo {volume} {18}},\
  \bibinfo {pages} {541} (\bibinfo {year} {2019})}\BibitemShut {NoStop}%
\bibitem [{\citenamefont {Sorella}\ \emph {et~al.}(2012)\citenamefont
  {Sorella}, \citenamefont {Otsuka},\ and\ \citenamefont
  {Yunoki}}]{Sorella2012}%
  \BibitemOpen
  \bibfield  {author} {\bibinfo {author} {\bibfnamefont {S.}~\bibnamefont
  {Sorella}}, \bibinfo {author} {\bibfnamefont {Y.}~\bibnamefont {Otsuka}}, \
  and\ \bibinfo {author} {\bibfnamefont {S.}~\bibnamefont {Yunoki}},\ }\href
  {\doibase 10.1038/srep00992} {\bibfield  {journal} {\bibinfo  {journal} {Sci.
  Rep.}\ }\textbf {\bibinfo {volume} {2}},\ \bibinfo {pages} {992} (\bibinfo
  {year} {2012})}\BibitemShut {NoStop}%
\end{thebibliography}%

\end{document}